\documentclass[12pt,preprint]{aastex}

\usepackage{graphicx}
\usepackage{epstopdf}

\newcommand{\kms}{km s$^{-1}$}
\newcommand{\ha}{H$\alpha$}
\newcommand{\solar}{\ifmmode_{\sun}\;\else$_{\sun}\;$\fi}

\newcommand{\HII}{H$\,${\sc ii}}
\newcommand{\HI}{H$\,${\sc i}}

\newcommand{\sighi}{$\Sigma_{\rm HI}$}
\newcommand{\sfr}{$\log {\rm SFR_D^{FUV}}$}

\newcommand{\dirr}{dIrr}
\newcommand{\rd}{$R_{\rm D}$}
\newcommand{\rbr}{$R_{\rm Br}$}
\newcommand{\rha}{$R_{\rm H\alpha}$}
\newcommand{\rfuvknot}{$R_{\rm FUV knot}$}
\newcommand{\cc}{$C_{coef}$}
\newcommand{\vdisp}{$v_{disp}$}
\newcommand{\sfra}{$\Sigma_{\rm SFR}$}

\begin{document}

\title{A Search for correlations between turbulence and star formation in LITTLE THINGS dwarf irregular galaxies}

\author{
Deidre A.\ Hunter\altaffilmark{1},
Bruce G.\ Elmegreen\altaffilmark{2},
Haylee Archer\altaffilmark{1},
Caroline E.\ Simpson\altaffilmark{3},
and Phil Cigan\altaffilmark{4}
}

\altaffiltext{1}{Lowell Observatory, 1400 West Mars Hill Road, Flagstaff, Arizona 86001, USA}
\altaffiltext{2}{IBM T\. J.\ Watson Research Center, 1101 Kitchawan Road, Yorktown Heights, New York USA}
\altaffiltext{3}{Department of Physics, Florida International University, CP 204, 11200 SW 8th St, Miami, Florida 33199 USA}
\altaffiltext{4}{George Mason University, 4400 University Dr., Fairfax, VA 22030-4444, USA}

\begin{abstract}
Turbulence has the potential for creating gas density enhancements that initiate
cloud and star formation (SF), and it can be generated locally by SF. To study
the connection between turbulence and SF, we looked for relationships between SF
traced by FUV images, and gas turbulence traced by kinetic energy density (KED)
and velocity dispersion (\vdisp) in the LITTLE THINGS sample of nearby \dirr\
galaxies. We performed 2D cross-correlations between FUV and KED images,
measured cross-correlations in annuli to produce correlation coefficients as a
function of radius, and determined the cumulative distribution function of the
cross correlation value. We also plotted on a pixel-by-pixel basis the locally
excess KED, \vdisp, and \HI\ mass surface density, \sighi, as determined from
the respective values with the radial profiles subtracted, versus the excess SF
rate density \sfra, for all regions with positive excess \sfra. We found that
\sfra\ and KED are poorly correlated. The excess KED associated with SF implies
a $\sim0.5$\% efficiency for supernova energy to pump local \HI\
turbulence on the scale of resolution here, which is a factor of $\sim2$
too small for all of the turbulence on a galactic scale. The excess \vdisp\ in
SF regions is also small, only $\sim0.37$ km s$^{-1}$. The local excess in
\sighi\ corresponding to an excess in \sfra\ is consistent with an \HI\
consumption time of $\sim1.6$ Gyr in the inner parts of the galaxies. The
similarity between this timescale and the consumption time for CO implies that
CO-dark molecular gas has comparable mass to \HI\ in the inner disks.
\end{abstract}

\keywords{galaxies: irregular --- galaxies: star formation --- galaxies: ISM --- galaxies: kinematics and dynamics}

\section{Introduction} \label{sec-intro}

The gas in the inner parts of spiral galaxies
is gravitationally unstable to the formation of clouds that can go on to form stars \citep{toomre,kennicutt89}.
However, in dwarf irregular (\dirr) galaxies, the atomic gas densities are much lower than in spirals and are apparently
stable against this instability \citep{sextansa,meurer96,vanzee97,hunter11}.
Furthermore, in inner spiral disks star formation increases as the gas density increases \citep{bigiel08},
while in dwarfs and outer spiral disks the atomic gas density cannot predict star formation rates \citep[SFRs,][]{bigiel10}.
So, what drives star formation in \dirr\ galaxies?

One process for creating clouds is compression of gas in a supersonically turbulent
medium \citep{elmegreen93,maclow04}. There is extensive evidence for interstellar
turbulence in galaxies, and turbulence in typical dIrrs has been shown to be
transonic \citep{burkhart10,maier17} while that in spirals is generally supersonic
\citep{maier16}. Furthermore, various distributions in stellar, cluster, and cloud
properties in dwarfs are consistent with sampling a fractal turbulent gas, including
composite cumulative \HII\ region luminosity functions
\citep{youngblood99,kingsburgh98}, stellar disk power spectra \citep{willett05},
mass functions of clouds and star clusters \citep{elmegreen97,hunter03,maclow04}, \ha\
probability distribution functions \citep{he04}, and the correlation between region
size and the star formation time scale \citep{efremov98}. \citet{dib05} found
evidence for scales in the interstellar medium (ISM) of Holmberg II  less than 6 kpc
in size that they interpret as due to a turbulence driver acting on that scale.
And, \citet{zhang12} showed from \HI\ spatial power spectra that either non-stellar
power sources are playing a fundamental role in driving the ISM turbulence or the
nonlinear development of turbulent structures has little to do with the driving
sources. In addition, \citet{hunter01,hunter11} have found regions of high velocity
dispersion in the \HI\ distribution of some \dirr\ galaxies that correlate with a
deficit of \HI\  in a manner suggestive of long-range, turbulent pressure
equilibrium \citep{piontek05}.

Turbulence can create density enhancements that initiate cloud formation \citep{krumholz05},
but turbulence also heats gas, which can make it harder to form clouds \citep{struck99}.
So, how important is turbulence in driving star formation in dwarfs?
It could be essential in outer disks where gas self-gravity is weak \citep{eh06}.
Also, a transition from subsonic to supersonic turbulence in the ISM could
be the cause of the transition in the Schmidt-Kennicutt star formation rate-gas density relationship
from inefficient star formation at low gas surface densities to star formation at higher densities \citep{kraljic14}.

Conversely, how important is star formation in driving turbulence? Simulations
suggest that stellar feedback and supernovae drive turbulence on the scale of the
galaxy thickness \citep{joung09,kim15}, and it may drive turbulence in molecular
clouds \citep{padoan16}, along with cloud self-gravity \citep{maclow17,ibanez17}.
Feedback destroys molecular clouds as well \citep{kim18}. Models also suggest
feedback controls the SFR by adjusting the disk thickness and midplane density
\citep{ostriker10} or by compressing nearby clouds, causing them to collapse
\citep{deharveng12,palmeirim17,egorov17}. On a galactic scale, feedback and
self-gravity operate together to drive turbulence
\citep[e.g.,][]{goldbaum16,krumholz18}. These models are uncertain, however.
Other simulations show no need for star formation to drive turbulence because they
reproduce the velocity dispersion from self-gravity alone; the only thing local
feedback needs to do is destroy the clouds where young stars form, preventing the
SFR from getting too large \citep{bournaud10,combes12,hopkins11}.

Observations are usually not decisive about the connection between the
SFR and turbulence. In a study of local dwarfs and low mass spirals,
\cite{stilp13} found a correlation between the core velocity dispersion in \HI\ line
profiles and the \HI\ surface density, suggestive of driving by gravitational
instabilities, but they also found a correlation with SFR at
$\Sigma_{\rm SFR}>10^{-4}\;M_\odot$ yr$^{-1}$ kpc$^{-2}$. \cite{stilp13} show that
both the \HI\ velocity dispersion and \sfra\ decrease with radius in a
galaxy; that makes correlations between these quantities ambiguous as they both
could depend on another parameter that varies with radius and not each other.

\cite{zhou17} studied 8 local galaxies with resolved spectroscopy and showed on a
pixel level that the velocity dispersion of ionized gas does not change over a
factor of $\sim40$ in SFR per unit area. Also for several hundred
local galaxies in the same survey, \cite{varidel20} found a very small correlation
between the galaxy-average vertical velocity dispersion of ionized gas and the total
SFR, with the dispersion increasing by only 6 km s$^{-1}$ for SFRs
between $10^{-3}$ and $10\;M_\odot$ yr$^{-1}$. This contrasts with
observations of high redshift galaxies where these authors show strong increases in
dispersion with SFR density and total rate, respectively, for rate
densities larger than $\sim0.1\;M_\odot$ yr$^{-1}$ kpc$^{-2}$ and rates larger than
$\sim3\;M_\odot$ yr$^{-1}$. This high-redshift correlation was earlier studied by
several groups, including \cite{lehnert13} who observed that the velocity dispersion
of ionized gas increases as the square root of the SFR per unit
area. \cite{lehnert13} concluded that star formation was the main driver of
turbulence and that it was also sufficient to maintain marginal stability in a disk.
On the other hand, \cite{ubler19} interpreted the increase in the ionized gas
velocity dispersion with SFR density for high redshift galaxies as
the result of gravitational instabilities, following the theory in
\cite{krumholz18}.

\cite{bacchini20} consider radial profiles of turbulent speeds and SFRs
in local spiral galaxies and account for all of the gas turbulence using
supernovae from young massive stars. They get more effective turbulence driving than
other studies because they include the radial increase in disk thickness, which
decreases the dissipation rate.

In this paper we look for evidence of a spatial correlation between star
formation and turbulence in the LITTLE THINGS sample of nearby dIrr galaxies. A
spatial correlation could be either a cause of star formation through the
production of a gas cloud or a result of star formation through mechanical energy
input to the local ISM through feedback from stars. We construct turbulent
Kinetic Energy Density (KED) maps from the kinetic energy associated with the bulk motions of the gas
- velocity from \HI\ velocity dispersion (moment 2) and mass from integrated column density (moment 0) maps,
per unit area in the galaxy.
We cross-correlate the KED maps with
far-ultraviolet (FUV) images that trace star formation over the past 200 Myr.
Because we are using intensity-weighted velocity dispersions, the ``turbulence''
includes all bulk motions of the gas, including thermal and turbulent. This follows
the two-dimensional (2D) cross-correlation method used by Ioannis Bagetakos (private communication) in
analysis of the spiral galaxy NGC 2403.

We also isolate turbulence in the vicinity of a SF region and determine the
excess KED and velocity dispersion from that region alone. This method removes any
background turbulence that may be generated by other means, such as gravitational
instabilities and collapse.

\section{Data} \label{sec-data}

LITTLE THINGS\footnote[5]{Funded in part by the National Science Foundation through grants AST-0707563, AST-0707426, AST-
0707468, and AST-0707835 to US-based LITTLE THINGS team members and with generous technical and
logistical support from the National Radio Astronomy Observatory.}
is a multi-wavelength survey of nearby dwarf galaxies \citep{lt12}.
The LITTLE THINGS sample is comprised of 37 \dirr\ galaxies and
4 Blue Compact Dwarf (BCD) galaxies. The galaxies are relatively nearby ($\leq$10.3 Mpc;
6\arcsec\ is $\leq$300 pc), contain gas so they have the potential for star formation,
and are not companions to larger galaxies. The sample also covers
a large range in dwarf galactic properties such as SFR and absolute magnitude.

We obtained \HI\ observations of the LITTLE THINGS galaxies with the
National Science Foundation's Karl G.\ Jansky Very Large Array
(VLA\footnote[6]{The VLA is a facility of the National Radio Astronomy Observatory.
The National Radio Astronomy Observatory is a facility of the National Science Foundation operated under cooperative agreement
by Associated Universities, Inc.}).
The \HI-line data are characterized by high sensitivity ($\leq1.1$
mJy beam$^{-1}$ per channel), high spectral resolution (1.3 or 2.6 \kms),
and high angular resolution (typically 6\arcsec).

Ancillary data used here include far-ultraviolet (FUV) images
obtained with the NASA {\it Galaxy Evolution Explorer} satellite
\citep[{\it GALEX\footnote[7]{{\it GALEX} was operated
for NASA by the California Institute of Technology under NASA contract NAS5-98034.}};][]{galex}.
These images trace star formation over the past 200 Myr.
These data also yield integrated SFRs \citep{hunter10} and the
radius at which we found the furthest out FUV knot \rfuvknot\ in each galaxy \citep{outerfuv}.
The SFRs are normalized to the area within one disk scale length \rd, although star formation is usually found beyond 1\rd.
\rd\ is measured from $V$-band surface brightness profiles \citep{herrmann13}.
Several of the LITTLE THINGS galaxies without {\it GALEX} FUV images are not included in this study (DDO 155, DDO 165, IC 10, UGC 8508).
Pixel values of FUV and \sfra\ are not corrected for extinction due to dust, which tends to be low in
these low metallicity galaxies.

The galaxy sample and characteristics that we use here are given in Table \ref{tab-gal}.
In some plots, we distinguish between those dIrrs that are classified as Magellanic irregulars (dIm)
and those that are classified as BCDs (Haro 29, Haro 36, Mrk 178, VIIZw403).

\begin{deluxetable}{lcccccccc}
\tabletypesize{\tiny}
\tablecaption{The Galaxy Sample \label{tab-gal}}
\tablewidth{0pt}
\tablehead{
\colhead{} & \colhead{D\tablenotemark{a}} & \colhead{M$_V$} & \colhead{$R_{\rm H\alpha}$\tablenotemark{b}}
& \colhead{$R_{\rm FUV knot}$\tablenotemark{c}}
& \colhead{$R_{\rm D}$\tablenotemark{d}}  & \colhead{$R_{\rm Br}$\tablenotemark{e}}
& \colhead{\sfr\tablenotemark{f}} \\
\colhead{Galaxy} &  \colhead{(Mpc)} & \colhead{} & \colhead{(kpc)}  & \colhead{(kpc)} & \colhead{(kpc)} & \colhead{(kpc)}
& \colhead{(M\solar\ yr$^{-1}$ kpc$^{-2}$)} & \colhead{} \\
}
\startdata
CVnIdwA  &  $3.6\pm0.08$  & $-12.37\pm0.09$ &      0.69 & 0.49$\pm$0.03 & 0.25$\pm$0.12 & 0.56$\pm$0.49 &   $-1.77\pm0.04$  \\
DDO 43    &  $7.8\pm0.8$  & $-15.06\pm0.22$ &      2.36 & 1.93$\pm$0.08 & 0.87$\pm$0.10 & 1.46$\pm$0.53 &     $-2.20\pm0.04$ \\
DDO 46    &  $6.1\pm0.4$  & $-14.67\pm0.16$ &      1.51 & 3.02$\pm$0.06 & 1.13$\pm$0.05 & 1.27$\pm$0.18 &     $-2.45\pm0.04$  \\
DDO 47    &  $5.2\pm0.6$  & $-15.46\pm0.24$ &      5.58 & 5.58$\pm$0.05 & 1.34$\pm$0.05 & \nodata             &     $-2.38\pm0.04$ \\
DDO 50    &  $3.4\pm0.05$  & $-16.61\pm0.03$ & \nodata & 4.86$\pm$0.03 & 1.48$\pm$0.06 & 2.65$\pm$0.27 &   $-1.81\pm0.04$  \\
DDO 52    & $10.3\pm0.8$ & $-15.45\pm0.17$ &      3.69 & 3.39$\pm$0.10 & 1.26$\pm$0.04 & 2.80$\pm$1.35 &     $-2.53\pm0.04$ \\
DDO 53    &   $3.6\pm0.05$ & $-13.84\pm0.03$ &      1.25 & 1.19$\pm$0.03 & 0.47$\pm$0.01 & 0.62$\pm$0.09 &   $-1.96\pm0.04$  \\
DDO 63    &   $3.9\pm0.05$ & $-14.78\pm0.03$ &      2.26 & 2.89$\pm$0.04 & 0.68$\pm$0.01 & 1.31$\pm$0.10 &   $-2.05\pm0.04$  \\
DDO 69    &   $0.8\pm0.04$ & $-11.67\pm0.11$ &      0.76 & 0.76$\pm$0.01 & 0.19$\pm$0.01 & 0.27$\pm$0.05 &   $-2.22\pm0.04$  \\
DDO 70    &   $1.3\pm0.07$ & $-14.10\pm0.12$ &      1.23 & 1.34$\pm$0.01 & 0.44$\pm$0.01 & 0.13$\pm$0.07 &  $-2.17\pm0.04$ \\
DDO 75    &   $1.3\pm0.05$ & $-13.91\pm0.08$ &      1.17 & 1.38$\pm$0.01 & 0.18$\pm$0.01 & 0.71$\pm$0.08 &  $-0.99\pm0.04$  \\
DDO 87    &   $7.7\pm0.5$ & $-14.98\pm0.15$ &      3.18 & 4.23$\pm$0.07 & 1.21$\pm$0.02 & 0.99$\pm$0.11 &  $-2.61\pm0.04$  \\
DDO 101  &   $6.4\pm0.5$ & $-15.01\pm0.16$ &      1.23 & 1.23$\pm$0.06 & 0.97$\pm$0.06 & 1.16$\pm$0.11 &  $-2.84\pm0.04$  \\
DDO 126  &   $4.9\pm0.5$ & $-14.85\pm0.24$ &      2.84 & 3.37$\pm$0.05 & 0.84$\pm$0.13 & 0.60$\pm$0.05 &   $-2.18\pm0.04$ \\
DDO 133  &  $3.5\pm0.2$ & $-14.75\pm0.16$ &      2.60 & 2.20$\pm$0.03 & 1.22$\pm$0.04 & 2.25$\pm$0.24 &     $-2.60\pm0.04$  \\
DDO 154  &   $3.7\pm0.3$ & $-14.19\pm0.16$ &      1.73 & 2.65$\pm$0.04 & 0.48$\pm$0.02 & 0.62$\pm$0.09 &   $-1.77\pm0.04$  \\
DDO 167   &  $4.2\pm0.5$ & $-12.98\pm0.25$ &      0.81 & 0.70$\pm$0.04 & 0.22$\pm$0.01 & 0.56$\pm$0.11 &   $-1.59\pm0.04$  \\
DDO 168   &  $4.3\pm0.5$ & $-15.72\pm0.25$ &      2.24 & 2.25$\pm$0.04 & 0.83$\pm$0.01 & 0.72$\pm$0.07 &   $-2.06\pm0.04$ \\
DDO 187   &  $2.2\pm0.07$ & $-12.68\pm0.07$ &      0.30 & 0.42$\pm$0.02 & 0.37$\pm$0.06 & 0.28$\pm$0.05 & $-2.60\pm0.04$  \\
DDO 210   &  $0.9\pm0.04$ & $-10.88\pm0.10$ & \nodata & 0.29$\pm$0.01 & 0.16$\pm$0.01 & \nodata             & $-2.66\pm0.04$ \\
DDO 216   &  $1.1\pm0.05$ & $-13.72\pm0.10$ &      0.42 & 0.59$\pm$0.01 & 0.52$\pm$0.01 & 1.77$\pm$0.45 & $-3.17\pm0.04$  \\
F564-V3    &  $8.7\pm0.7$ & $-13.97\pm0.18$ & \nodata & 1.24$\pm$0.08 & 0.63$\pm$0.09 & 0.73$\pm$0.40 &  $-2.94\pm0.04$ \\
IC 1613     &  $0.7\pm0.05$ & $-14.60\pm0.16$ & \nodata & 1.77$\pm$0.01 & 0.53$\pm$0.02 & 0.71$\pm$0.12 &  $-1.97\pm0.04$  \\
LGS 3        &  $0.7\pm0.08$ &  $-9.74\pm0.25$ & \nodata & 0.32$\pm$0.01 & 0.16$\pm$0.01 & 0.27$\pm$0.08 &  $-3.75\pm0.04$ \\
M81dwA     & $3.6\pm0.2$ & $-11.73\pm0.13$ & \nodata & 0.71$\pm$0.03 & 0.27$\pm$0.00 & 0.38$\pm$0.03 &  $-2.30\pm0.04$  \\
NGC 1569  & $3.4\pm0.2$ & $-18.24\pm0.13$ & \nodata & 1.14$\pm$0.03 & 0.46$\pm$0.02 & 0.85$\pm$0.24 &  $-0.32\pm0.04$  \\
NGC 2366  & $3.4\pm0.3$ & $-16.79\pm0.20$ &      5.58 & 6.79$\pm$0.03 & 1.91$\pm$0.25 & 2.57$\pm$0.80 &  $-2.04\pm0.04$  \\
NGC 3738  & $4.9\pm0.5$ & $-17.12\pm0.24$ &      1.48 & 1.21$\pm$0.05 & 0.77$\pm$0.01 & 1.16$\pm$0.20 &   $-1.52\pm0.04$  \\
NGC 4163  & $2.9\pm0.04$ & $-14.45\pm0.03$ &      0.88 & 0.47$\pm$0.03 & 0.32$\pm$0.00 & 0.71$\pm$0.48 &  $-1.89\pm0.04$ \\
NGC 4214  & $3.0\pm0.05$ & $-17.63\pm0.04$ & \nodata & 5.46$\pm$0.03 & 0.75$\pm$0.01 & 0.83$\pm$0.14 &  $-1.11\pm0.04$ \\
Sag DIG     & $1.1\pm0.07$ & $-12.46\pm0.14$ &      0.51 & 0.65$\pm$0.01 & 0.32$\pm$0.05 & 0.57$\pm$0.14 &  $-2.40\pm0.04$  \\
WLM           & $1.0\pm0.07$ & $-14.39\pm0.15$ &      1.24 & 2.06$\pm$0.01 & 1.18$\pm$0.24 & 0.83$\pm$0.16 & $-2.78\pm0.04$  \\
Haro 29      &  $5.8\pm0.3$ & $-14.62\pm0.11$ &      0.96 & 0.86$\pm$0.06 & 0.33$\pm$0.00 & 1.15$\pm$0.26 &  $-1.21\pm0.04$ \\
Haro 36      &  $9.3\pm0.6$ & $-15.91\pm0.15$ &      1.06 & 1.79$\pm$0.09 & 1.01$\pm$0.00 & 1.16$\pm$0.13 &   $-1.88\pm0.04$ \\
Mrk 178      &  $3.9\pm0.5$ & $-14.12\pm0.26$ &      1.17 & 1.45$\pm$0.04 & 0.19$\pm$0.00 & 0.38$\pm$0.00 &   $-1.17\pm0.04$  \\
VIIZw403    &  $4.4\pm0.07$ & $-14.27\pm0.04$ &      1.27 & 0.33$\pm$0.04 & 0.53$\pm$0.02 & 1.02$\pm$0.29 &  $-1.80\pm0.04$ \\
\enddata
\tablenotetext{a}{Distance to the galaxy. References are given by \citet{lt12}.}
\tablenotetext{b}{Radius of furthest out detected \HII\ region \rha\ in each galaxy from \citet{he04}.
Galaxies without \HII\ regions or with \HII\ regions
extending beyond the area imaged do not have $R_{H\alpha}$.}
\tablenotetext{c}{Radius of furthest out detected FUV knot \rfuvknot\ in each galaxy from \citet{outerfuv}.
Galaxies without {\it GALEX} images have no value for this radius.}
\tablenotetext{d}{Disk scale length \rd\ determined from the $V$-band image surface photometry from \citet{herrmann13}.
In the case of galaxies with breaks in their surface brightness profiles, we have chosen the scale length that describes the
primary underlying stellar disk.}
\tablenotetext{e}{Break radius \rbr\ where the $V$-band surface brightness profile changes slope given by \citet{herrmann13}.
DDO 47 and DDO 210 do not have breaks in their surface brightness profiles.}
\tablenotetext{f}{SFR measured from the integrated FUV luminosity and normalized to the area within one \rd\ from \citet{hunter10}.
The normalization is independent of the radial extent of the FUV emission in a galaxy.}
\end{deluxetable}

\section{Cross-correlations} \label{sec-xc}

\subsection{Two-dimensional}

KED and FUV images were the inputs to the 2D cross-correlation. We geometrically
transformed the FUV image to match the orientation and field of view (FOV) of the
\HI\ map using \textsc{OHGEO} in the Astronomical Image Processing System
(AIPS) and then smoothed it to the \HI\ beam using \textsc{SMOTH} in AIPS.
We also blanked the pixels outside of the galaxy FUV emission, replacing the
blanked pixels with zeros, so that pure noise would not add to the correlation
coefficient \cc. We constructed the KED maps as  $0.5 \times N_{\rm HI} \times
v_{disp}^2$, where $N_{\rm HI}$ is the \HI\ column density in hydrogen atoms per
cm$^2$ and \vdisp\ is the velocity dispersion in km s$^{-1}$. The conversion from
counts in the KED maps to ergs pc$^{-2}$ is given for each galaxy in Table
\ref{tab-corr}. Prior to executing the cross-correlations, we scaled both the FUV
and KED images so that the pixel values were the same order of magnitude (roughly
100).  These KED values determined from \HI\ column density have not been
multiplied by 1.36 to include Helium and heavy elements.  This factor will be used
later when the efficiency of KED generation is calculated.

We decided not to remove the underlying exponential disks for the 2D cross-correlations.
Although the SFR drops off with radius, the FUV image consists of knots of young stars and there can be large
FUV knots in the outer disks.
For the \HI\ moment 0 and 2 maps, the \HI\ surface density and velocity dispersion do, on average, change with
radius too, but not in a regular and homogenous fashion.
Thus, in the 2D \cc\ maps exponential structure could remain.

Here,
a \cc\ of 1 is perfectly correlated such that every bump and wiggle in one map is exactly
reproduced in the other. A value of $-1$ is perfectly anti-correlated.
The amplitude of the peak is a measure of the
coincidence of features in each image. If the KED image correlates well with the
local FUV flux, then the peak will be high and the breadth will be the average size
of their rms summed feature sizes.

\begin{figure}[t!]
\epsscale{1.2}
\vskip -1.4truein
\plotone{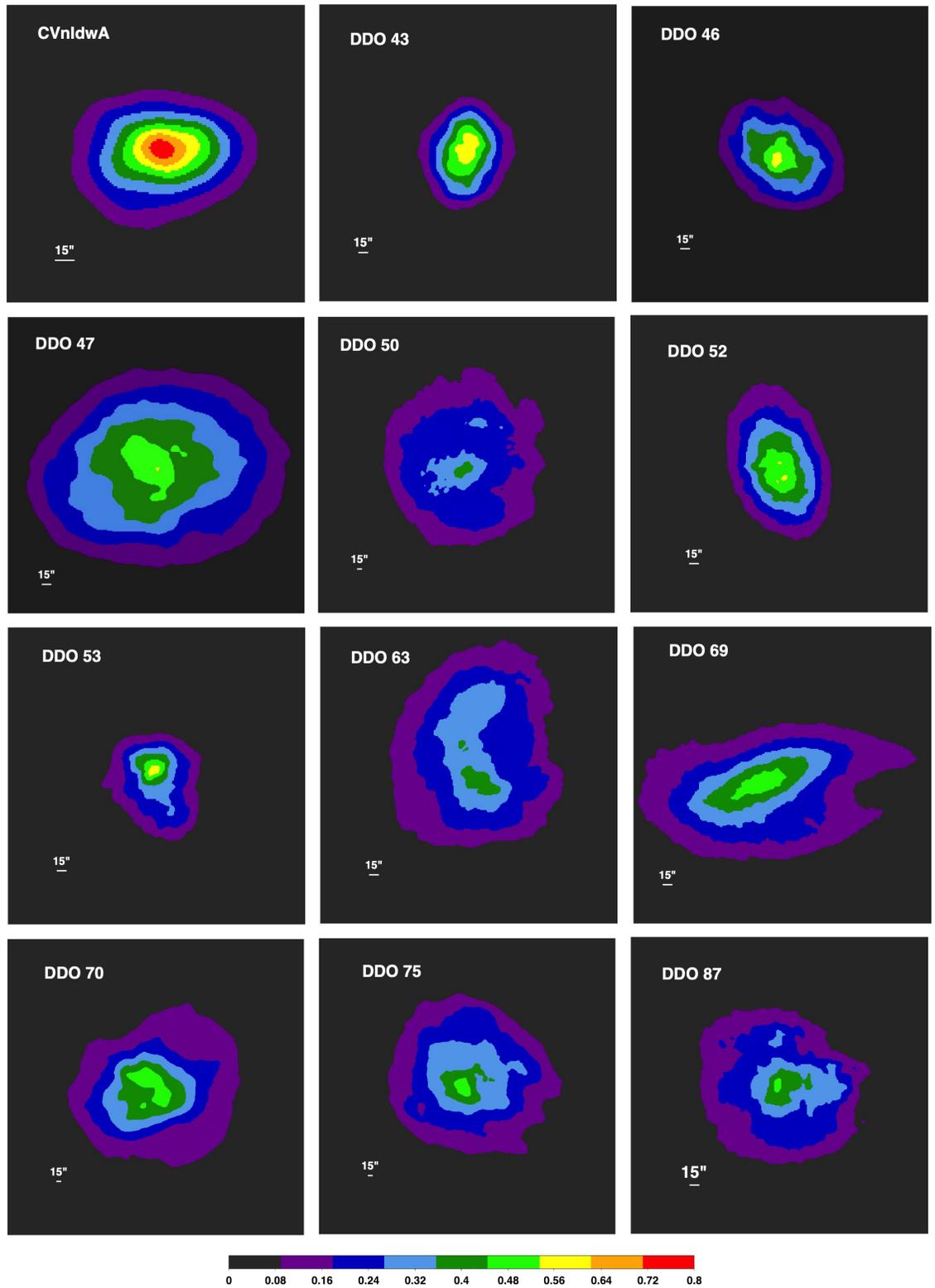}
\vskip -1.25truein
\caption{Cross-correlation matrices for each galaxy. The images are displayed with the same color scale from
\cc\ of zero to 0.8.
\label{fig-xc}}
\end{figure}

We used the commands \textsc{correl\_images} and \textsc{corrmat\_analyze} in IDL
with a python wrapper. We used this command to shift one image relative to the other
over and over again to yield a map of \cc. The peak pixel value in the \cc\ map is
the adopted \cc. For example, for NGC 2366, we did a 150$\times$150 array of
offsets. That is, we calculated the \cc\ for x,y offset of $-150$, $-150$ to x,y
offset of $+150$, +$150$. This produces a matrix of 301$\times$301 pixels. The peak,
in this case, is at pixel 145, 147 and has a value of 0.3 compared to the center
pixel 151, 151 value of 0.26.
Thus, the maximum correlation is achieved when the FUV image is shifted
relative to the KED image by the offset corresponding to x,y of 145, 147.
We checked the \cc\ of a piece of one of the galaxies
``by hand'' with a Fortran program we wrote and we obtained the same peak \cc.
The peak \cc\ and x,y shifts to that pixel are
given in Table \ref{tab-corr}. The cross-correlation matrices are shown in Figure
\ref{fig-xc}.

The shift in x,y is also given in Table \ref{tab-corr} relative to the disk scale-length \rd,
for better comparison to the size of the galaxy. The shifts vary between 0.02\rd\ (IC 1613)
and 4.75\rd\ (Haro 36). 50\% of the galaxies (18) have shifts less than 0.5\rd,
33\% (12) have shifts between 0.5\rd\ and 1\rd, and 17\% (6) have shifts greater than 1\rd.

\includegraphics[scale=0.85]{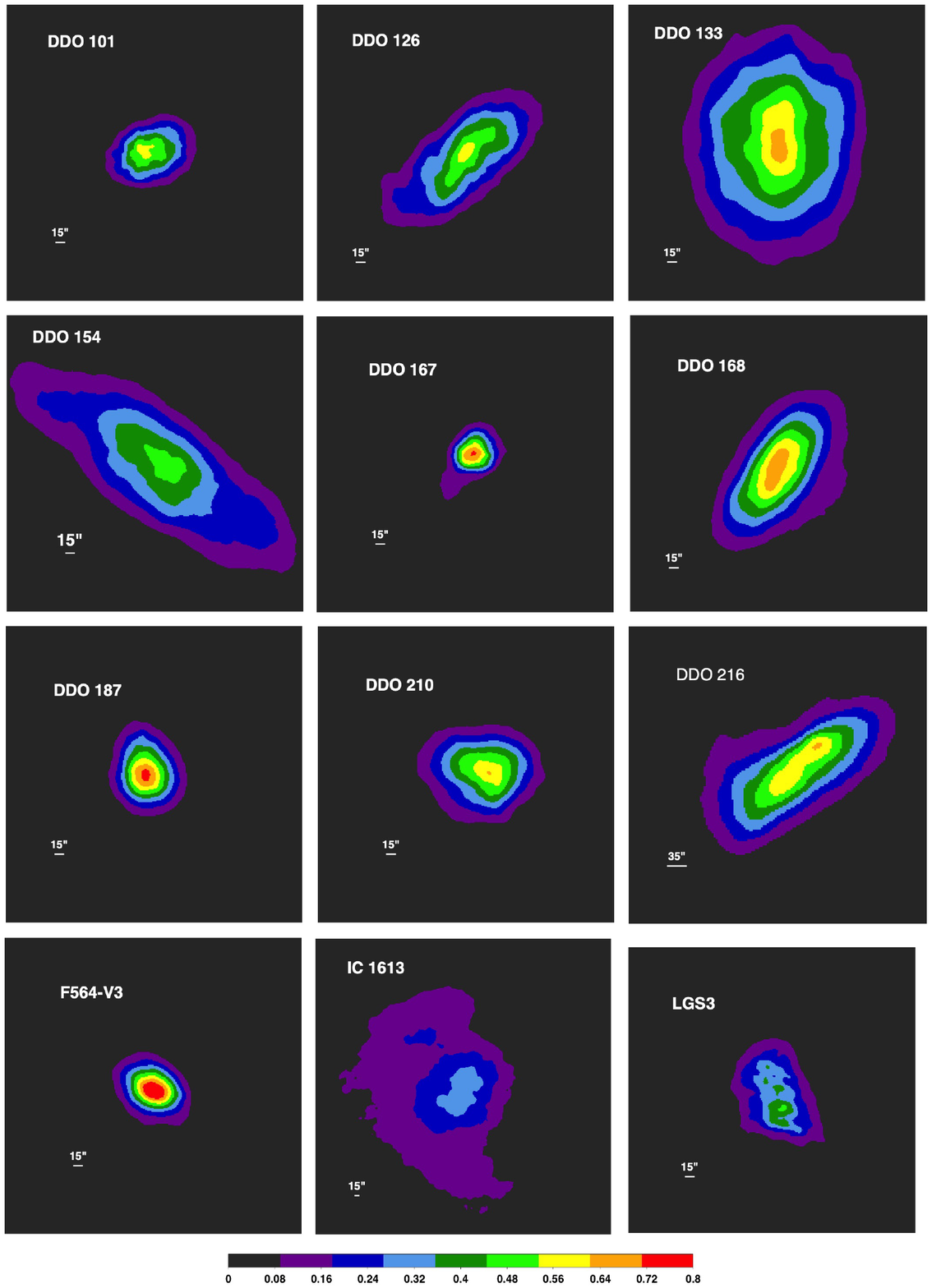}

\includegraphics[scale=0.85]{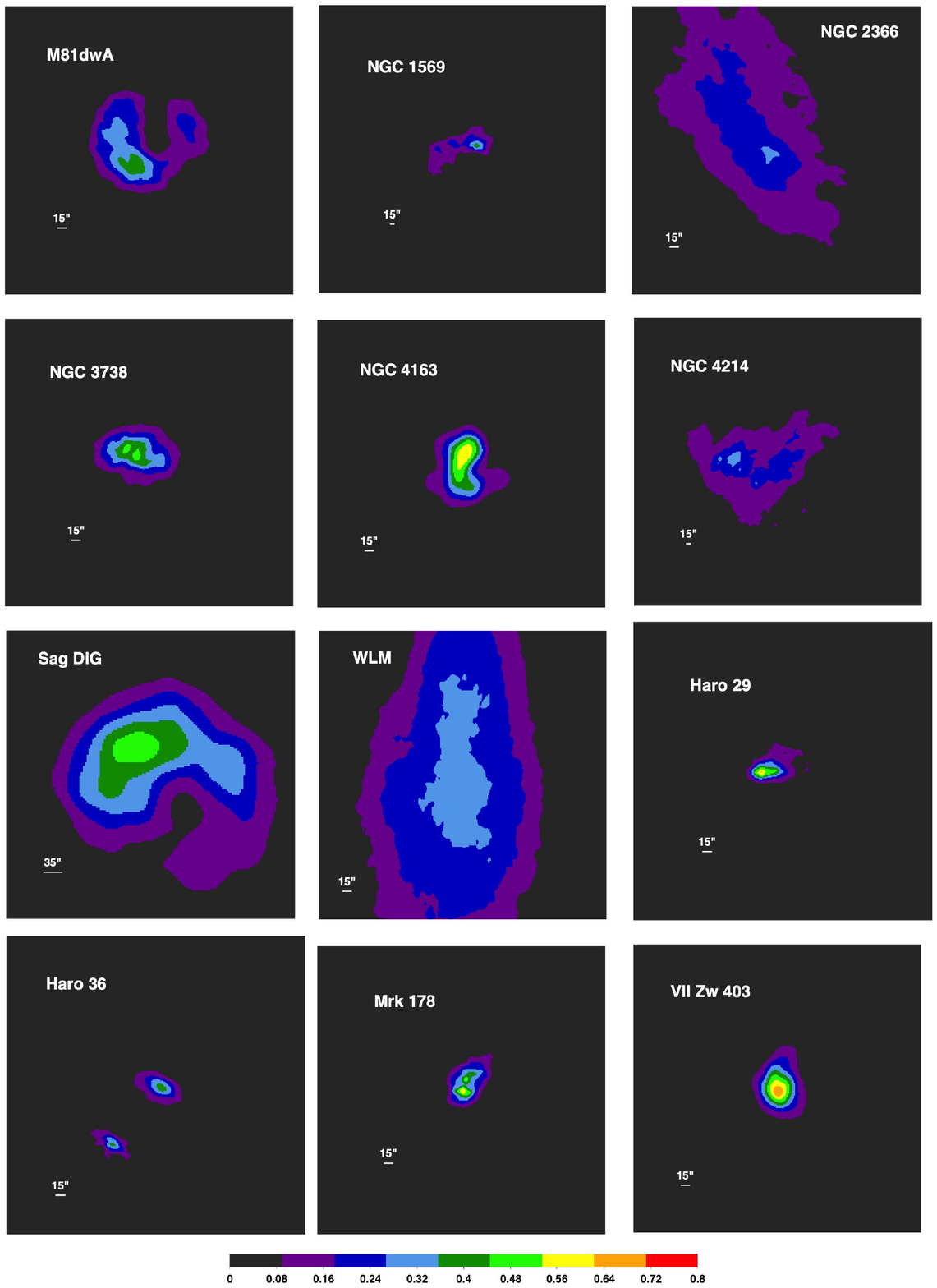}

Ioannis Bagetakos (private communication) examined the cross-correlation method on NGC 2403 as a function of
image scale, focussing on scales of 0.23 to 3 kpc, and found correlations on various scales for different images
such as star formation tracers, dust, and \HI.
Thus, we divided our images up into square sub-regions 16$\times$16, 32$\times$32, 64$\times$64, and 128$\times$128 pixels
and computed the \cc\ in each box.
The coefficient images constructed from this just look like noise and show no particular connection to the FUV image.
So we do not consider them further.

\begin{figure}[t]
\epsscale{0.8}
\vskip -0.3truein
\plotone{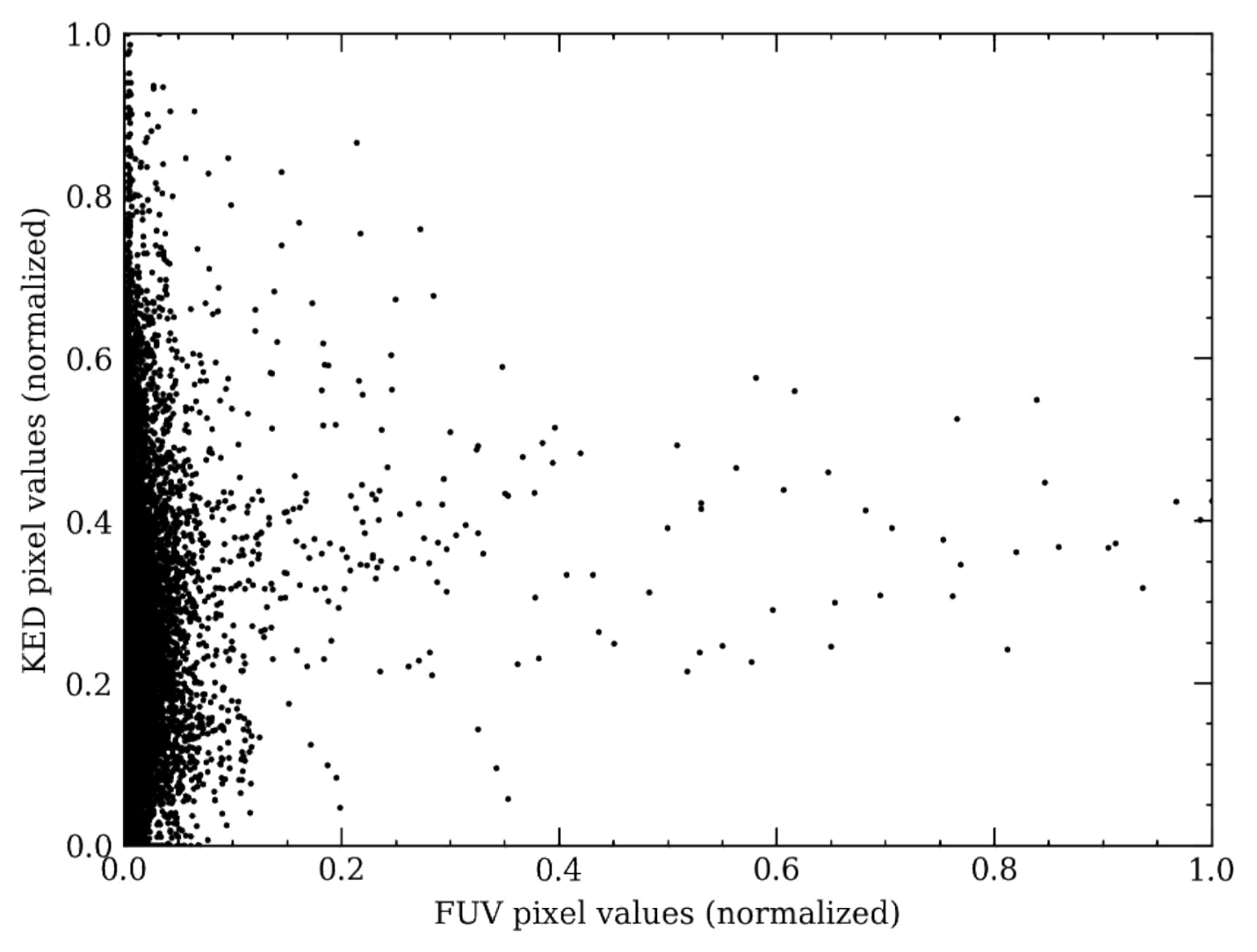}
\vskip -0.25truein
\caption{Values of pixels in KED image plotted against values of pixels in FUV image for NGC 2366.
The images have each been normalized so that pixel values are between 0 and 1.
A notable correlation would be expected to appear as a cluster of points in the top right corner
(or, in the other corners near values of 1.0 for anti-correlation), however this evidence is not seen.
\label{fig-pix-pix}}
\end{figure}

We also applied alternate methods on one galaxy, NGC 2366 with a max \cc\ of 0.3, to
examine the robustness of our approach.
This galaxy was chosen for initial and special tests because it has a giant \HII\ region and the
\HI\ velocity dispersion is high around this region, making it an interesting candidate for looking for
a star formation-turbulence correlation.
One problem with cross-correlations, in particular, can be caused by
moderate signal-to-noise (S/N) pixels dampening the value of \cc. One simple diagnostic is a plot of
the pixel values of the KED image against the pixel values of the FUV image, given
that the FUV image has been geometrically transformed and smoothed to the same pixel grid and beam size as the
KED image. We normalized the pixel values in each image to range from 0 to 1, and
this plot is shown in Figure \ref{fig-pix-pix}. If there were a notable correlation,
we would expect a cluster of points in the top right corner. If the images were
anti-correlated, we would expect clusters of points around the top left and bottom
right. We do not see either of these extremes.
While there are some points in the top left corner, it is not a distinct cluster,
rather it appears to be consistent with a typical tail end of a simple distribution of points from 0 to 1.

We also tried variations of weighted normalized cross-correlations and a wavelet analysis to NGC 2366.
The zero-mean normalized cross correlation coefficient (ZNCC) is basically the standard Pearson Correlation measure $\rho$ for a 2D image.
Applied to NGC 2366, ZNCC is 0.26.  Like $\rho$ or $r$ coefficients, $+1$ is perfect correlation, $-1$ is perfect anti-correlation,
and so 0.26, which is what we found for the max \cc,  implies a not very significant degree of correlation.
One way to deal with pixels with low S/N is to use a weighted normalized cross-correlation (WNCC).
For this test, we weighted the pixels by the ratio of their signal to the standard deviation of values in the map,
which is effective at down-weighting background noise pixels.
Using this method, we obtain a WNCC value of -0.023 -- effectively zero, implying no significant correlation between the two images.

\begin{figure}[t!]
\epsscale{0.7}
\vskip -0.3truein
\plotone{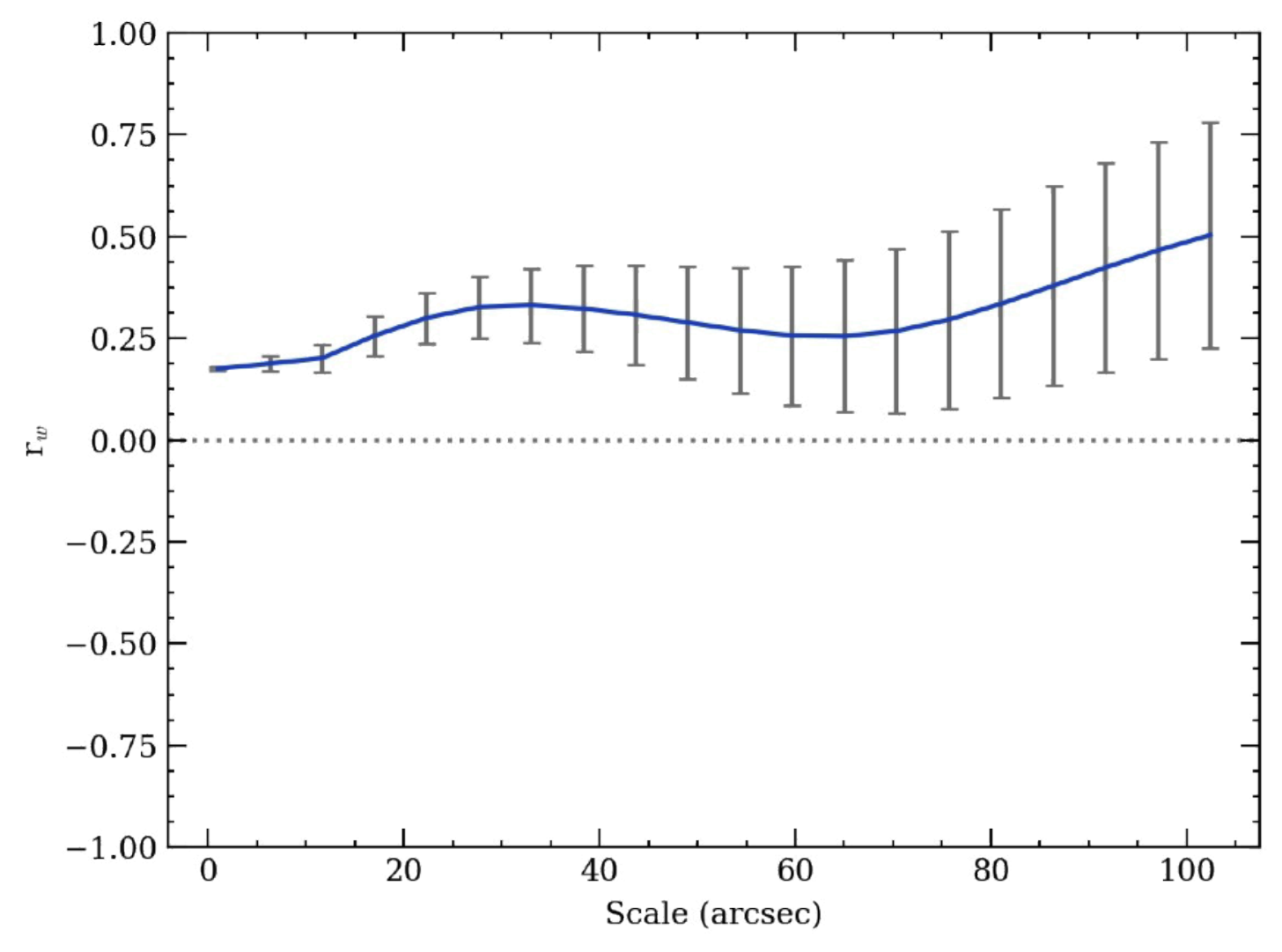}
\vskip -0.2truein
\caption{Cross-correlation coefficient $r_w$ for NGC 2366 KED and FUV images convolved with progressively larger `Mexican hat' kernels.
We find no significant correlation between the two images at any resolved scale.
\label{fig-wavelet}}
\end{figure}

For our final test on NGC 2366, we used a wavelet analysis to see if the degree of correlation depends on scale/resolution.
In this process, each image is convolved with progressively larger 2D kernels or wavelets, in this case a Ricker or `Mexican Hat' wavelet,
and the cross-correlation is calculated at each of these scales or 'lags'.
The result for NGC 2366 is shown in Figure \ref{fig-wavelet}.
Strong correlations at a particular spatial scale would be evidenced by wavelet cross correlation $r_w$ values of $\gtrsim$ 0.6 at that scale.
Eventually, as the images are convolved to large enough scales, they become less resolved and therefore naturally correlate.
Figure \ref{fig-wavelet} indicates that there is not much difference at any of the resolved scales.
Using a slightly modified wavelet from the literature \citep[e.g.,][]{ossenkopf08},
there may be evidence of a slightly more prominent correlation between the images at 30 pixel scales (45\arcsec),
but $r_w$ is still not significant.

Thus, we conclude that no matter how we look at the FUV and KED images of NGC 2366,
the two images are mildly correlated at best and this does not change much with
scale.

\begin{figure}[t!]
\epsscale{0.4}
\vskip -1truein
\plotone{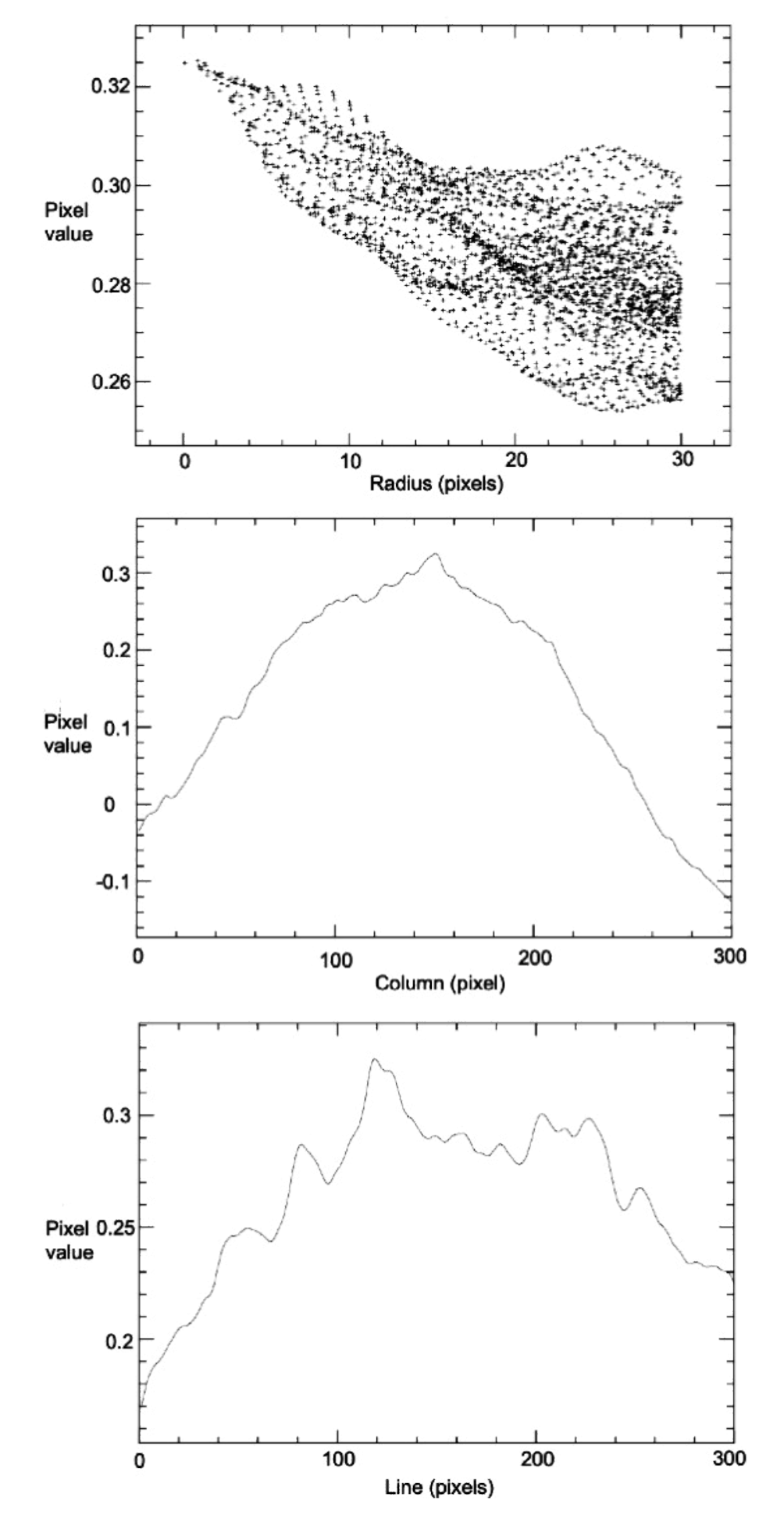}
\vskip -0.2truein
\caption{Cuts through the peak in the \cc\ matrix of WLM.
Top: Radial profile.
Middle: Row plot.
Bottom: Column plot.
\label{fig-wlmcuts}}
\end{figure}

The width of the peak signal in a cross-correlation matrix is expected to represent the scale of the
correlation. However, in our matrices, the width is not well defined. The issue is demonstrated
in Figure \ref{fig-wlmcuts} where we show a radial plot and row and column cuts through the
peak of the WLM \cc\ matrix.
The peak is, of course, obvious, but the radial plot is messy and the single row and column cuts
show a complex background.
The main feature in the cross correlation maps is the exponential disk because both the KED and the SFR density
peak in the center with the exponential disk.
The width of the \cc\ in Figure \ref{fig-wlmcuts} is influenced more by the width of the disk than the scale of the 2D correlation.
What to take as the baseline for a fit to the peak is also not clear.
Therefore, we do not consider the widths of the peaks further here.

\begin{deluxetable}{lcccccc}
\tabletypesize{\tiny}
\tablecaption{Correlation Coefficients and Offsets \label{tab-corr}}
\tablewidth{0pt}
\tablehead{
\colhead{Galaxy} &  \colhead{Max \cc} & \colhead{X shift\tablenotemark{a}} & \colhead{Y shift\tablenotemark{a}}
& Shift/\rd\ & \colhead{Offset (X$\times$Y)\tablenotemark{b}} & \colhead{Calibration ($10^{41}$)\tablenotemark{c}} \\
}
\startdata
CVnIdwA         &    0.77      &       3    &    2      &   0.38   & 75$\times$75       &  7.67  \\
DDO 43            &   0.61       &      4    &   13      &  0.89   & 150$\times$150    &  18.33   \\
DDO 46            &   0.57       &      -5   &    -9      &  0.40   & 150$\times$150     &  26.77  \\
DDO 47           &    0.54        &      2   &    -4       &  0.13   & 150$\times$150     &  9.32  \\
DDO 50           &    0.41       &      -2   &    -8       &  0.14   & 300$\times$300    &  20.40   \\
DDO 52            &   0.56       &       6   &  -14       &  0.91  & 150$\times$150    &  24.86   \\
DDO 53            &   0.58       &      -4   &     5       &  0.36  & 150$\times$150    &   24.54  \\
DDO 63           &    0.41       &        8   &   -5       &  0.39 & 150$\times$150     &  18.81  \\
DDO 69           &    0.50      &      -15   &   -9      &   0.54  & 150$\times$150    &  28.29   \\
DDO 70            &   0.50      &      -23   &  18      &  0.63   & 300$\times$300    &   4.83  \\
DDO 75            &  0.52       &        -8   &   -2      &  0.43   & 300$\times$300    &  18.09   \\
DDO 87          &    0.48       &       -2    &   2       &  0.09  & 150$\times$150     &  18.73  \\
DDO 101         &   0.59       &      -15   &    5      &  0.76   & 150$\times$150    &  15.22   \\
DDO 126         &   0.62        &         2  &     3     &  0.15    & 150$\times$150   &  22.95    \\
DDO 133          &  0.65       &         1    &   2      &  0.05    & 150$\times$150     &  6.61  \\
DDO 154         &   0.50       &       10    &  -4      &  0.60  & 150$\times$150     &  17.69  \\
DDO 167        &    0.72        &        9    &  11      &  1.97   & 150$\times$150    &  22.95   \\
DDO 168        &    0.68        &       -2   &    0      &  0.08  & 150$\times$150     &  19.36  \\
DDO 187        &    0.74       &        -8    &   0      &  0.35   & 150$\times$150     & 25.58   \\
DDO 210        &    0.63       &       23   &    1       &  0.94  & 150$\times$150     &  8.77  \\
DDO 216        &    0.63       &       20   &  15       &  0.90  &  75$\times$75       &  3.54  \\
F564-V3          &     0.87      &         0   &  -4        & 0.40 & 150$\times$150     &  8.69   \\
IC 1613            &   0.33       &        1   &    2        &  0.02  & 300$\times$300    &  17.69  \\
LGS 3              &   0.46       &      13   & -19        &  0.73 & 150$\times$150     &  8.05  \\
M81dwA          &   0.42       &    -11  &  -16         &  1.88 & 150$\times$150      &  18.01  \\
NGC  1569      &   0.40       &     30   &    6          &  1.65 & 300$\times$300     &  29.08   \\
NGC  2366      &  0.30       &      -6   &    -4         &  0.09 & 150$\times$150      &  21.51  \\
NGC  3738      &  0.48       &    -13   &     7         & 0.68 & 150$\times$150      &  25.50  \\
NGC  4163      &  0.62       &       4   &   12         & 0.83  & 150$\times$150     &   15.46  \\
NGC  4214     &   0.35       &    -88   &    8         & 2.57 & 300$\times$300      &  18.17  \\
SagDIG          &  0.51       &      -7   &  15        &  0.97  & 75$\times$75          &  1.84   \\
WLM              &   0.33       &      0  &  -32         & 0.20 & 150$\times$150        &  22.95  \\
Haro 29          &  0.63       &    -21 &    -1         & 2.69 & 150$\times$150       &  23.19   \\
Haro 36          &  0.34        &   -46  & -54          & 4.75 & 150$\times$150       &  21.91  \\
Mrk 178          &  0.60       &        2 &    -1         & 0.33 & 150$\times$150       &  25.98  \\
VIIZw 403       &   0.68        &       1  &   -3          & 0.19 & 150$\times$150      &   12.11  \\
\enddata
\tablenotetext{a}{Offset of the pixel with the maximum \cc\ from the center of the
array, in pixels. The pixel scale is 1.5\arcsec\ except for DDO 216 and Sag DIG
where it is 3.5\arcsec.} \tablenotetext{b}{Offsets in pixels.}
\tablenotetext{c}{Constant by which to convert counts in KED maps to ergs
pc$^{-2}$.}
\end{deluxetable}

\subsection{Radial profiles}

\begin{figure}[t!]
\epsscale{1.0}
\vskip -1.0truein
\plotone{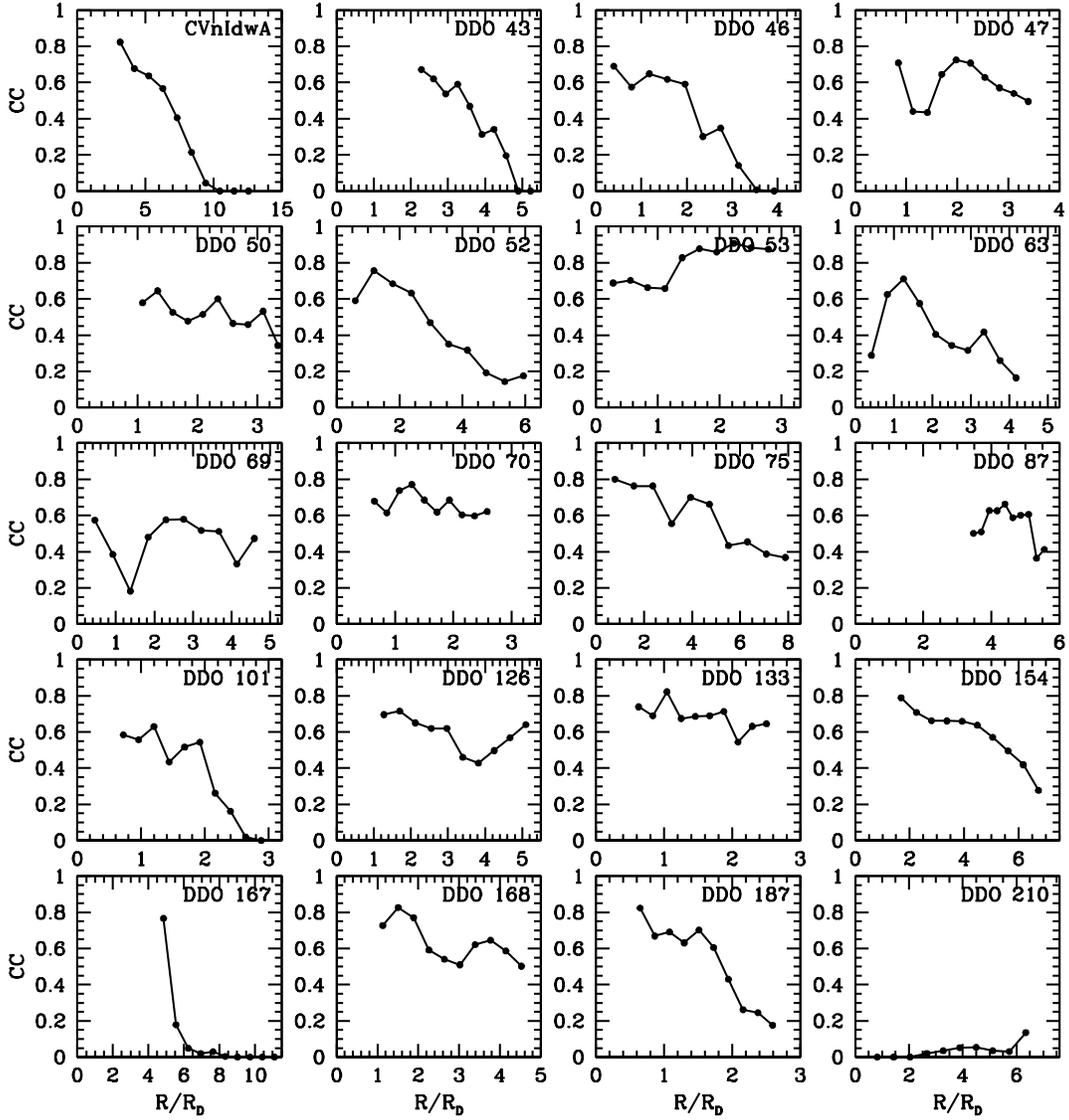}
\vskip -1.9truein
\caption{Correlation coefficient between FUV and KED images in annuli as a function of distance from the center of the galaxy.
The \cc\ profile is plotted from 0 to 1 for all galaxies for ease of comparison, and the radius is
normalized by the disk scale length measured from the $V$-band image (Table \ref{tab-gal}).
The pixel values in each annulus have been normalized by the average in the annulus, so large-scale trends with radius have been removed.
\label{fig-ccannuli}}
\end{figure}

We also calculated the \cc\ in annuli from the center of the galaxy outward.
The image was blanked outside of the target annulus, which were chosen to match
those used by \citet{lt12} to produce the \HI\ radial profiles.
We normalized the pixel values in the annulus with respect to the average
in the annulus, so in effect large-scale variations, such as the exponential fall-off with radius, are taken out.
Then we measured the \cc\ for the annulus.
Figure \ref{fig-ccannuli} shows the \cc\ of the annuli as a function of annulus distance from the center of the galaxy.
The annuli used galaxy center, ellipticity, and major axis position angle determined from $V$-band images
and given by \citet{lt12}.

We see a wide variety of profiles.
The central points in NGC 4163 and in VIIZw403 reach a \cc\ of nearly 0.95 and a few other galaxies have peaks as high as 0.9.
By contrast the peak in DDO 210 occurs in the outermost annulus and only reaches a value of 0.14.
In many galaxies the \cc\ drops in value with radius,
but in many others it is relatively flat.
In a few galaxies the \cc\ drops precipitously from a relatively high value for the inner-most annulus to near zero
beyond that radius (DDO 167, F564-V3, Haro 36).

\clearpage

\hspace{-0.85truein}
\vspace{-2.0truein}
\includegraphics[scale=0.9]{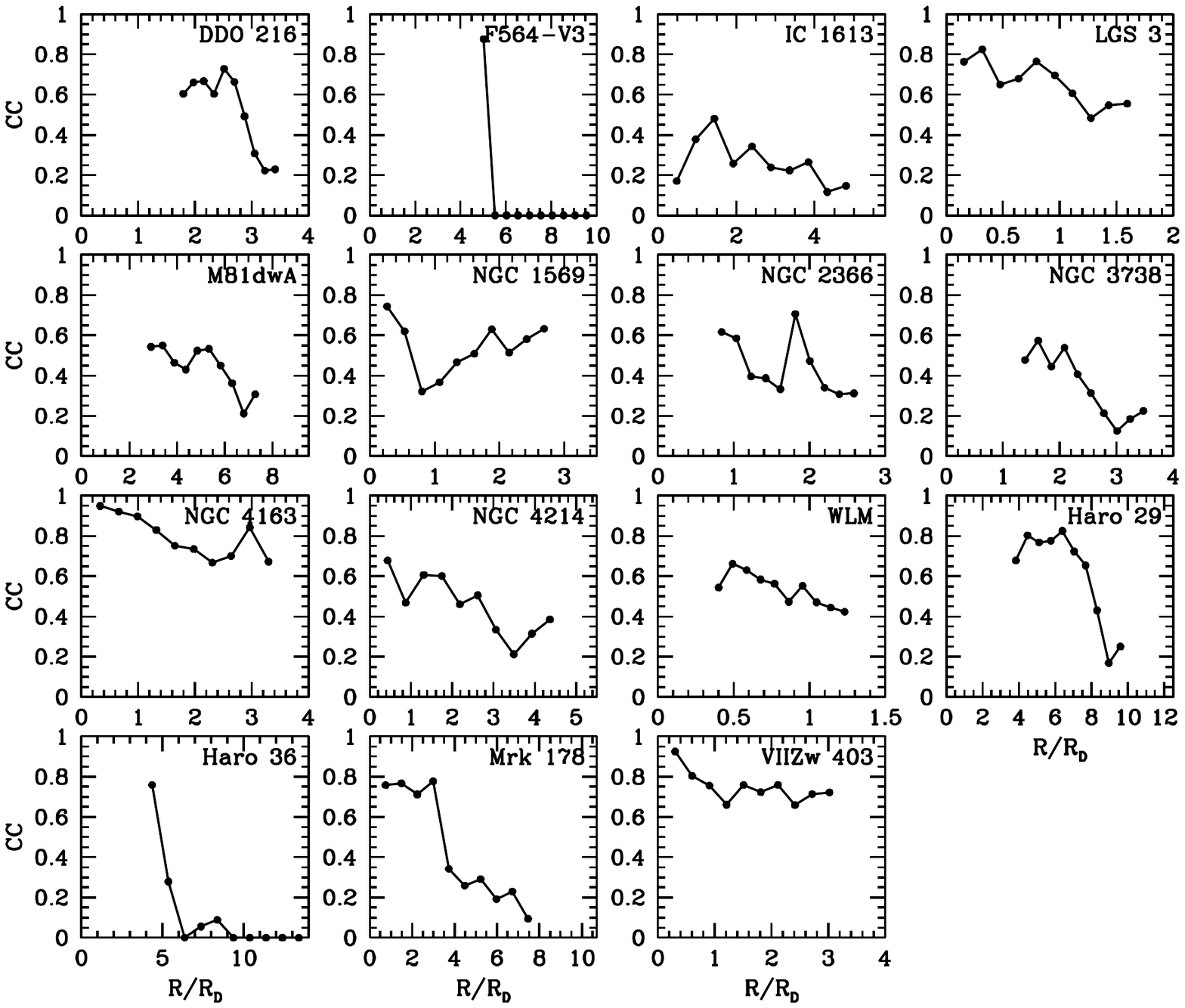}

\section{Results} \label{sec-results}

\subsection{Cross-correlations}

\begin{figure}[t!]
\epsscale{0.8}
\plotone{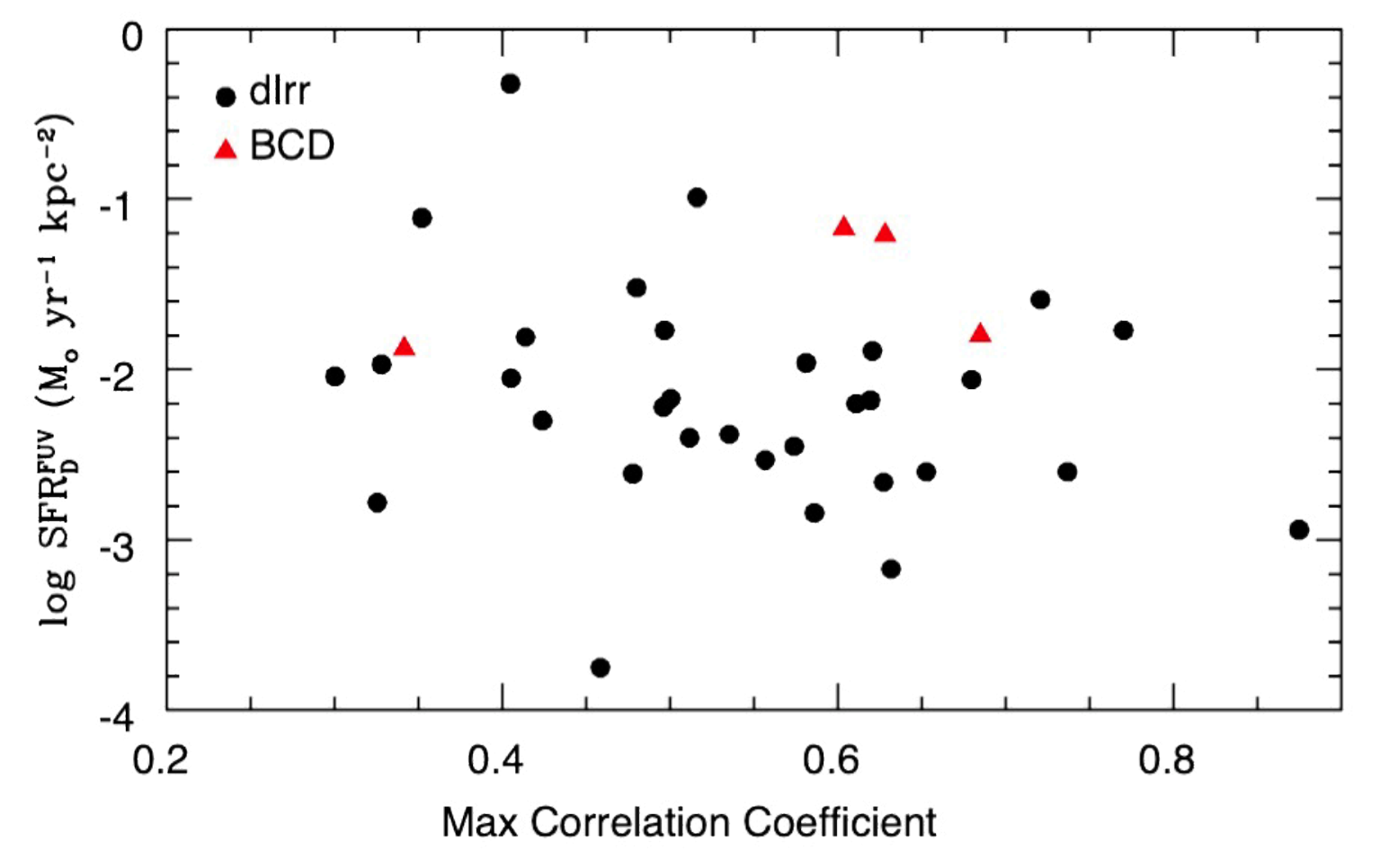}
\vskip -0.25truein
\caption{Integrated SFR normalized to one disk scale length versus the maximum correlation coefficient for each galaxy.
The correlation coefficients are given in Table \ref{tab-corr} and the SFRs are in Table \ref{tab-gal}.
\label{fig-sfr}}
\end{figure}

Generally, the 2D \cc\ indicate low levels of correlation between the FUV and KED
images. In Figure \ref{fig-sfr} we plot the peak \cc\ against the integrated SFR for
each galaxy to see if a higher level of correlation is related to the overall SFR.
There is no relationship between the two values. In annuli, \cc\ can be as high as
0.9 in the center, indicating a correlation, but the values tend to be low overall,
and the radial profiles exhibit a wide range of shapes.
From the images, visually most of the FUV is patchy and tends to be concentrated towards the central regions of the galaxies
while the \HI\ often extends quite far outside the optical/UV galaxy.
So the birds-eye view of a \dirr\ might expect a higher correlation in the central regions where
there is ample \HI\ and FUV, with little correlation as you go farther out where there are typically fewer FUV knots.

By comparison, in the spiral
NGC 2403 Ioannis Bagetakos (private communication) found that the FUV and \HI\ surface mass density are
uncorrelated with a \cc$<0.20$. They did, however, find correlations between dust
and star formation (\cc$>0.55$) and between PAHs and \HI\ (\cc$\sim0.55$).
Bagetakos et al.\ chose NGC 2403 as their pilot galaxy because it is in the THINGS sample \citep{walter08}
with \HI\ data, as well as images at 8 microns, 24 microns, \ha, and FUV, and is nearby with an
\HI\ beam of 136 pc $\times$ 119 pc. As an Scd spiral it is significantly larger and more massive than the \dirr\ galaxies
in this study.

\subsection{Degree of lumpiness}

\begin{figure}[t!]
\epsscale{1.0}
\vskip -1truein
\plotone{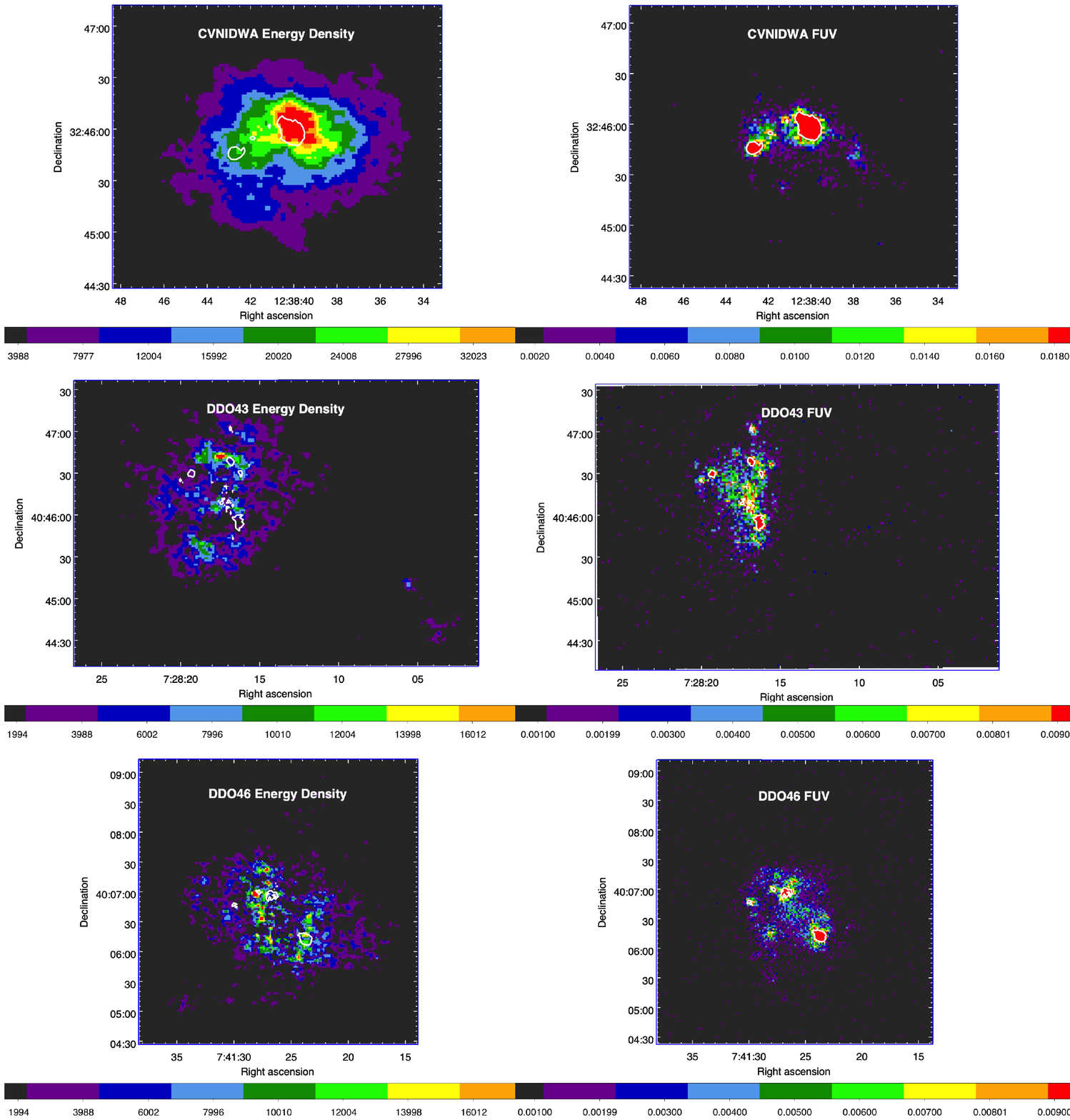}
\vskip -1truein
\caption{Kinetic energy density (KED) maps and full-resolution FUV images for each galaxy.
The major FUV knots are contoured and
that white contour is shown on the KED map to facilitate comparison.
The conversion of counts in the KED maps to physical units is given in Table \ref{tab-corr}.
To convert FUV counts s$^{-1}$ to flux in units of erg s$^{-1}$ cm$^{-2}$ \AA$^{-1}$
multiply by $1.4\times10^{-15}$.
Figures for the rest of the galaxies in this study are available in the on-line materials
(72 images in 12 figures).
\label{fig-edandfuv}}
\end{figure}

Since star formation is usually lumpy, we ask whether
the lack of correlation between FUV and KED images is because KED is smooth compared to FUV or because lumps
in the two images do not
correlate. Figure \ref{fig-edandfuv} shows the KED maps and FUV images at full
resolution. A contour of the FUV image is superposed on the KED map
to facilitate comparison.
We see that FUV and KED maps are both generally lumpy
although the lumps are not necessarily located in the same place.

\begin{figure}[t!]
\epsscale{1.0}
\hspace{-0.5truein}
\plotone{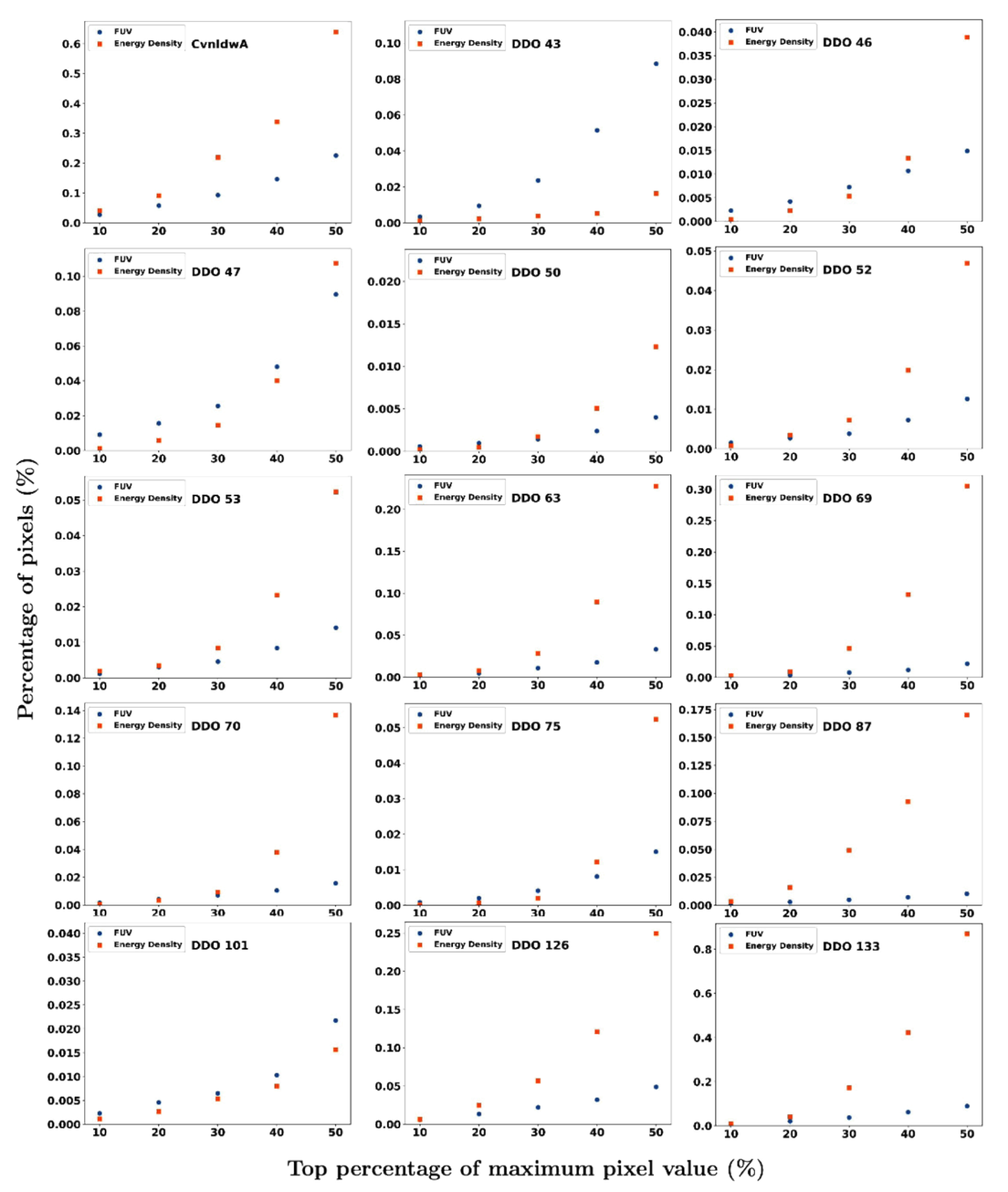}
\vspace{-0.2truein}
\caption{
Percentage of pixels with values above a given percentage of the maximum value for FUV and KED images.
\label{fig-fracpixels}}
\end{figure}

To examine the degree of lumpiness, we looked at the fraction of pixels with raw
values above a given percentage of the maximum pixel value in the image.
Specifically, we counted the fraction of total pixels that have counts within 10\%,
20\%, 30\%, 40\%, and 50\% of the maximum count value in each of the FUV and KED
images. These data are 
shown in Figure
\ref{fig-fracpixels} as percentage of total pixels as a function of selected cut-off
deviation from the maximum pixel value in the image. For example in CVnIdwA, the
percentage of pixels with values within 10\% of the maximum value is 0.69\% in the
FUV image and 1.03\% in the KED image, whereas the percentage of pixels with values
within 50\% of the maximum value is 5.65\% in the FUV image and 15.98\% in the KED
image.

To understand what these plots mean we can compare the appearance of the galaxies in
Figure \ref{fig-edandfuv} with the plots in Figure \ref{fig-fracpixels}. We see in
the images that galaxies like LGS3, DDO 87, DDO 133, and SagDIG have a few small FUV
knots but more or bigger KED knots. The KED knots fill more of the area and so
a higher fraction of the pixels are close to the peak intensity. These galaxies
have flat FUV profiles in Figure \ref{fig-fracpixels} because very few pixels
are close to the peak intensity, i.e., the FUV is spotty, but they have KED
profiles that rise with percentage of maximum pixel value because the KED is
more uniform. DDO 43 is unique in this sample because it is the only one with
an approximately flat KED profile and an FUV profile that rises with percentage of
maximum pixel value. The reason is clear from Figure \ref{fig-edandfuv}, which
shows that the FUV image of DDO 43 is filled with bright spots, making most of the
image close to the peak pixel value, while the KED image has weaker peaks that are
more spread out. DDO 167, on the other hand, has FUV and KED knots that
are comparable in size, and FUV and KED profiles that rise together with
percentage of maximum pixel value, as do DDO 47, DDO 101, F564-v3, and NGC 4163.
Most of the galaxies have broader KED distributions than FUV emission, so their KED
pixel percentages rise faster than their FUV pixel percentages as the top percentage
of the maximum pixel value increases.

\begin{figure}[h!]
\epsscale{0.5}
\hspace{-1.0truein}\plotone{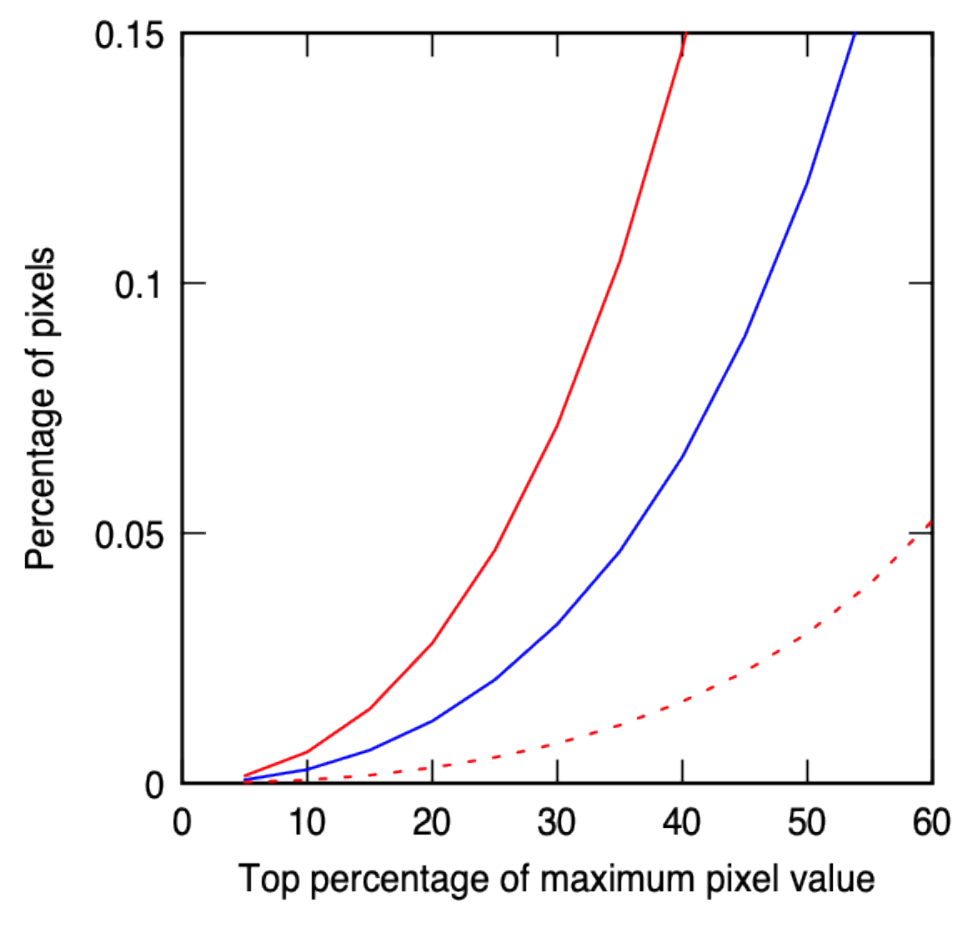}
\vspace{-0.25truein}
\caption{
Model for the curves in Figure \ref{fig-fracpixels} based on exponential profiles.
The top curve has a scale length 1.5 times larger than the middle curve and an overall galaxy
size the same, 2 scale lengths. The lower curve also has a scale length 1.5 times
larger than the middle curve but the overall galaxy size for the lower curve is 1.5 times larger.
Larger scale lengths for a given galaxy size
make the percentage curves rise faster because more of the disk is close to the
peak intensity at the center.
\label{haylee_fraction}}
\end{figure}

The general rising trend of the curves in Figure \ref{fig-fracpixels} is mostly
the result of the exponential radial profile of the disk with the peaks in KED and
FUV standing a nearly fixed fraction above the mean profile. Figure
\ref{haylee_fraction} shows models for these curves assuming an exponential disk
intensity profile $I(r)=e^{-r}$, so the radius as a function of intensity is
$r(I)=-\ln(I)$. The radius at 10\% of the peak is then $r(10\%)=-\ln(1-0.1)$, and
the number of pixels brighter than that is the area of the circle at this radius, or
$\pi r(10\%)^2$.  In general, for an intensity that is the fraction $x$ down from
the peak intensity, the fraction of pixels in the total disk is
\begin{equation}
f(x)=\pi \left(-\ln[1-x]\right)^2/\left(\pi r_{\rm max}^2\right)
\end{equation}
where $r_{\rm max}$ is the size of the disk measured in scale lengths. Figure
\ref{haylee_fraction} shows $f(x)$ versus $x$ in three cases. The top curve is for
an exponential profile with a scale length 1.5 times larger than the middle curve
and an overall galaxy size the same, $r_{\rm max}=2$ scale lengths. The lower curve
also has a scale length 1.5 times larger than the middle curve but the overall
galaxy size for the lower curve is 1.5 times larger ($r_{\rm max}=3$). Larger scale
lengths for a given galaxy size make the percentage curves rise faster because more
of the disk is close to the peak intensity at the center.

The similarity of the model curves in Figure \ref{haylee_fraction} to the
observations in Figure \ref{fig-fracpixels} implies that the qualitative effect
being captured by the fractional distribution is the result of the exponential disk.
However, the percentage of pixels observed is much smaller than the model
percentage, i.e., several percent or less for the observations compared to
$\sim10$\% at the 50\% top percentage of maximum pixel value. This difference
implies that the peaks in the KED and FUV distributions stand above the exponential
disk, so their areas are a small fraction, $\sim10$\%, of the disk area, but the
peak intensities have about the same radial dependence as the average disk, which
means they are a fixed factor times the average disk brightness.

\hspace{-1.1truein} \includegraphics[scale=0.95]{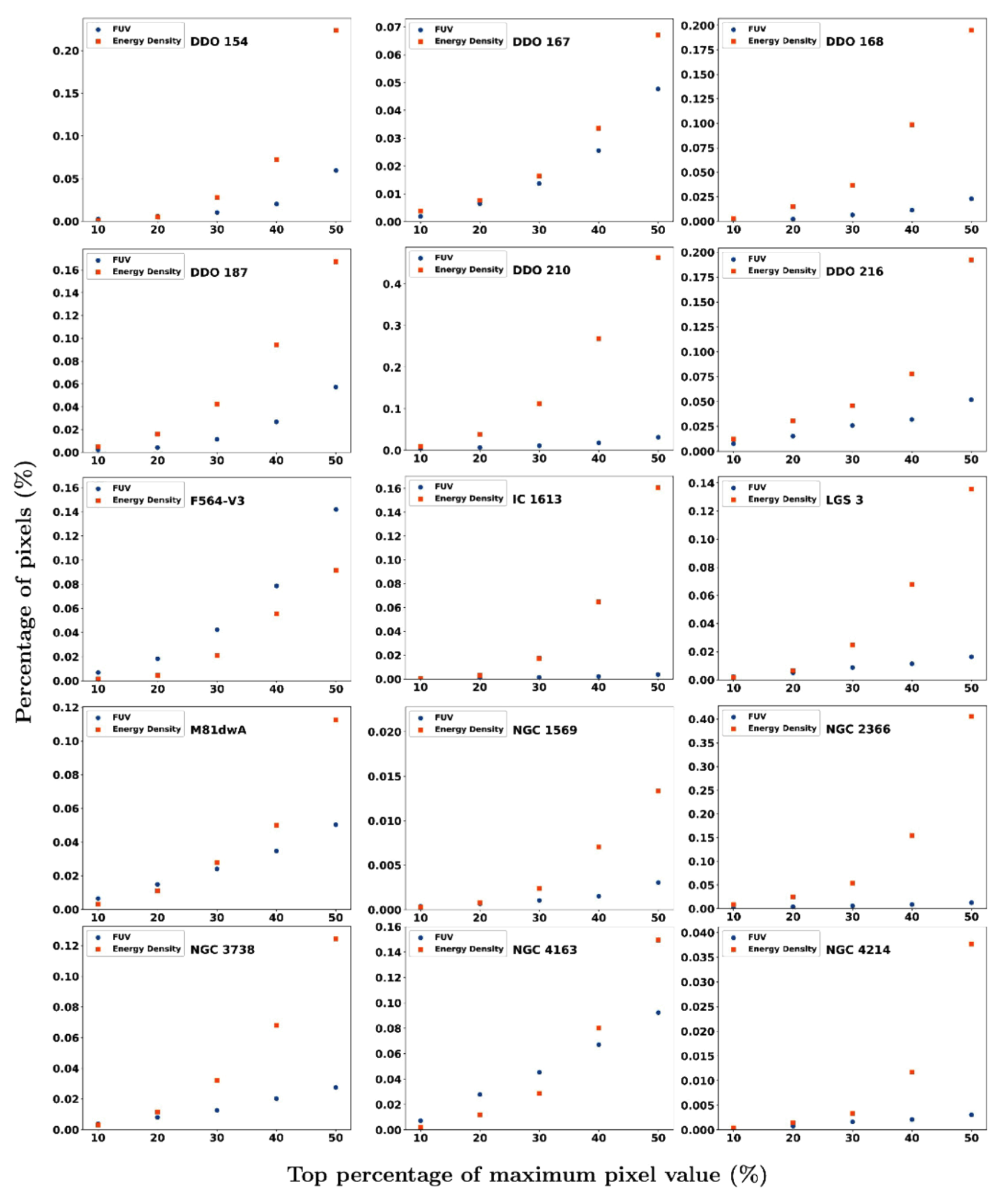}

\hspace{-1.1truein} \includegraphics[scale=0.95]{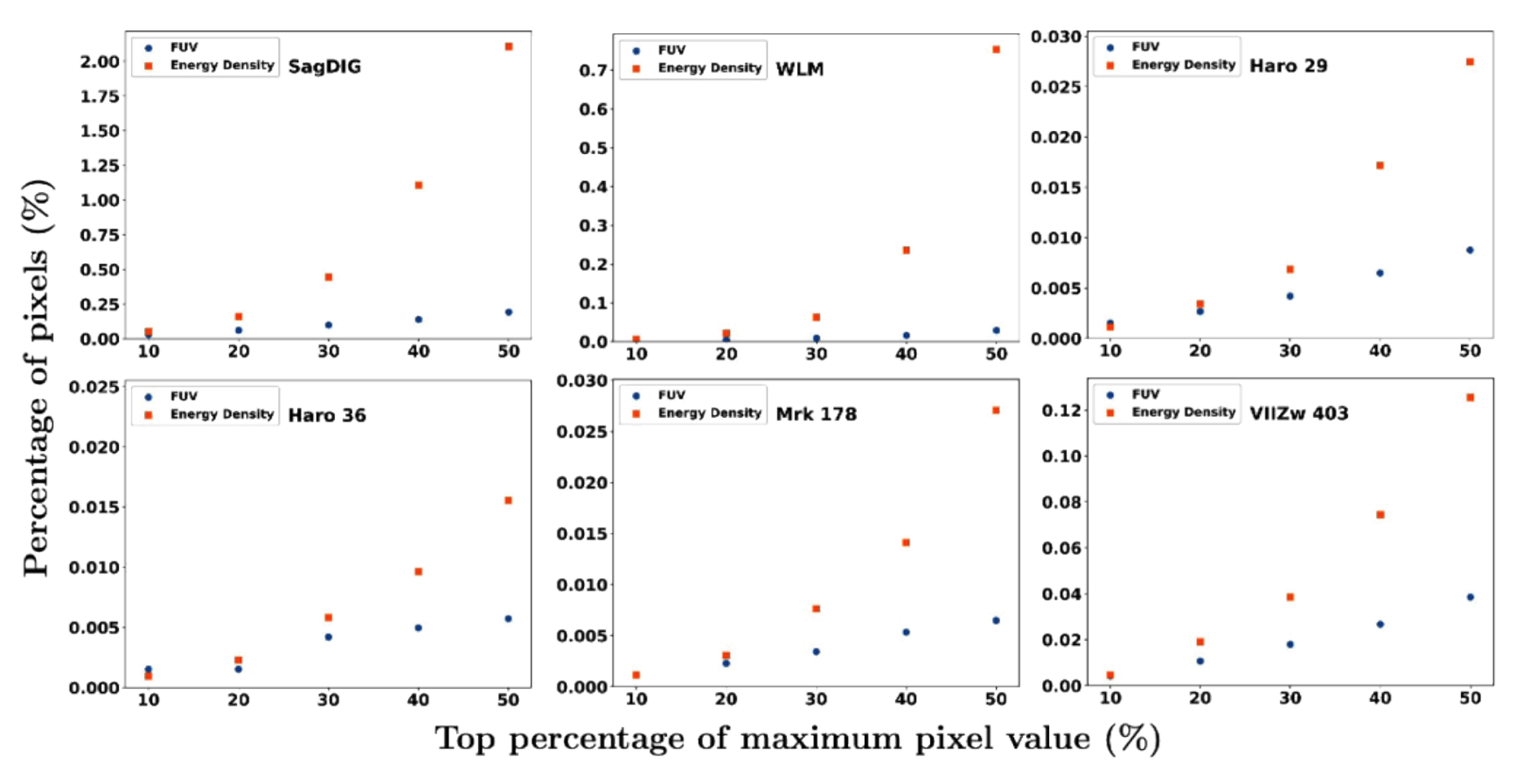}

\subsection{Pixel-pixel scatter plots}

\begin{figure}[t!]
\epsscale{1.0}
\vskip -0.5truein
\plotone{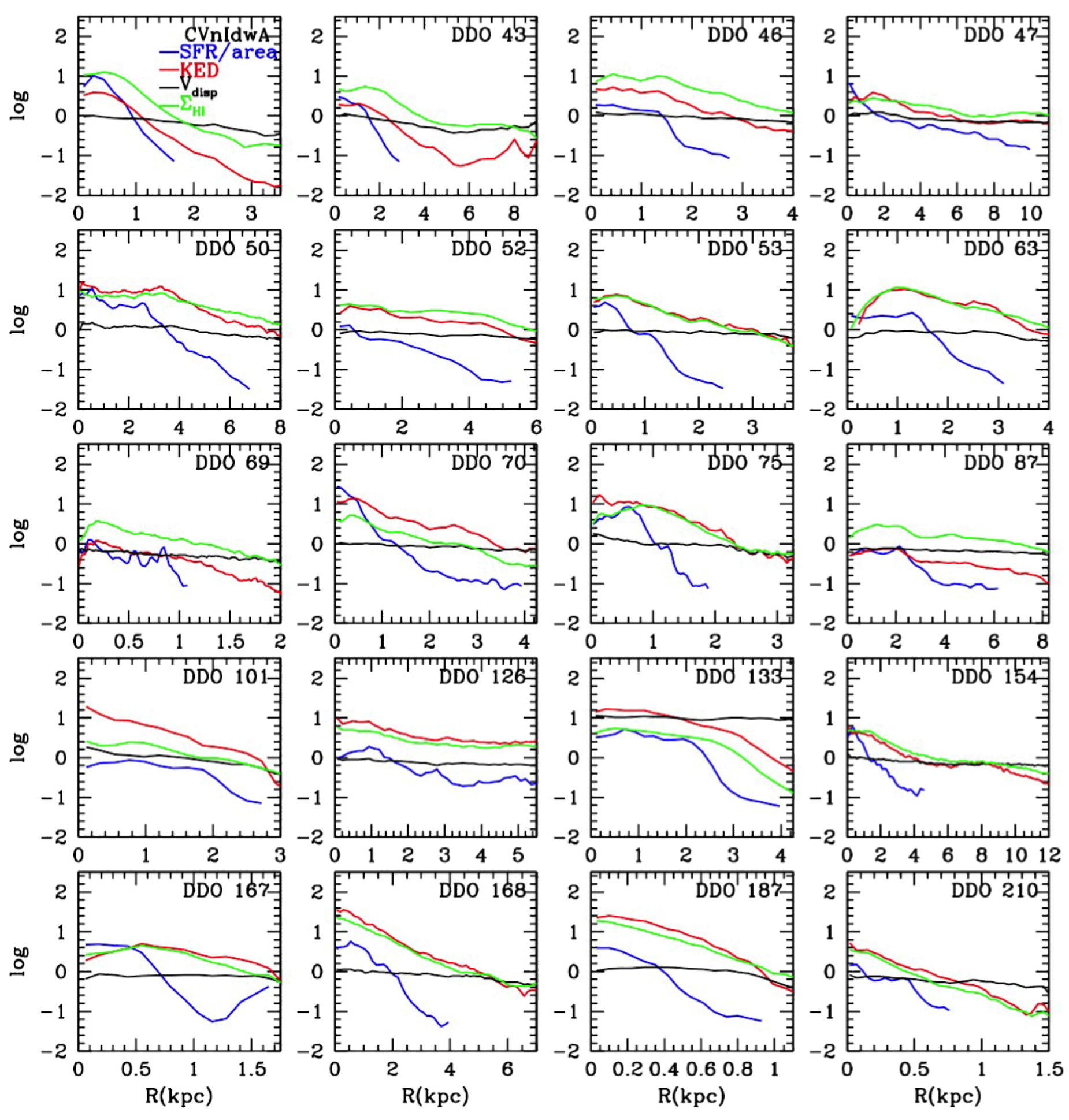}
\vskip -0.2truein
\caption{
Azimuthally-averaged radial profiles of \sfra\ determined from the FUV,
KED (not corrected for He and heavy elements),
\vdisp, and \sighi. FUV emission is the limiting quantity in that it
does not go out as far as the other quantities. Optical disk parameters (center, $b/a$, and major axis
position angle) from \citet{lt12} were used, and holes in the gas or FUV emission were not used
in the averages.
\label{fig-radialprof}}
\end{figure}

\begin{figure}[t!]
\epsscale{1.0}
\vskip -0.7truein
\plotone{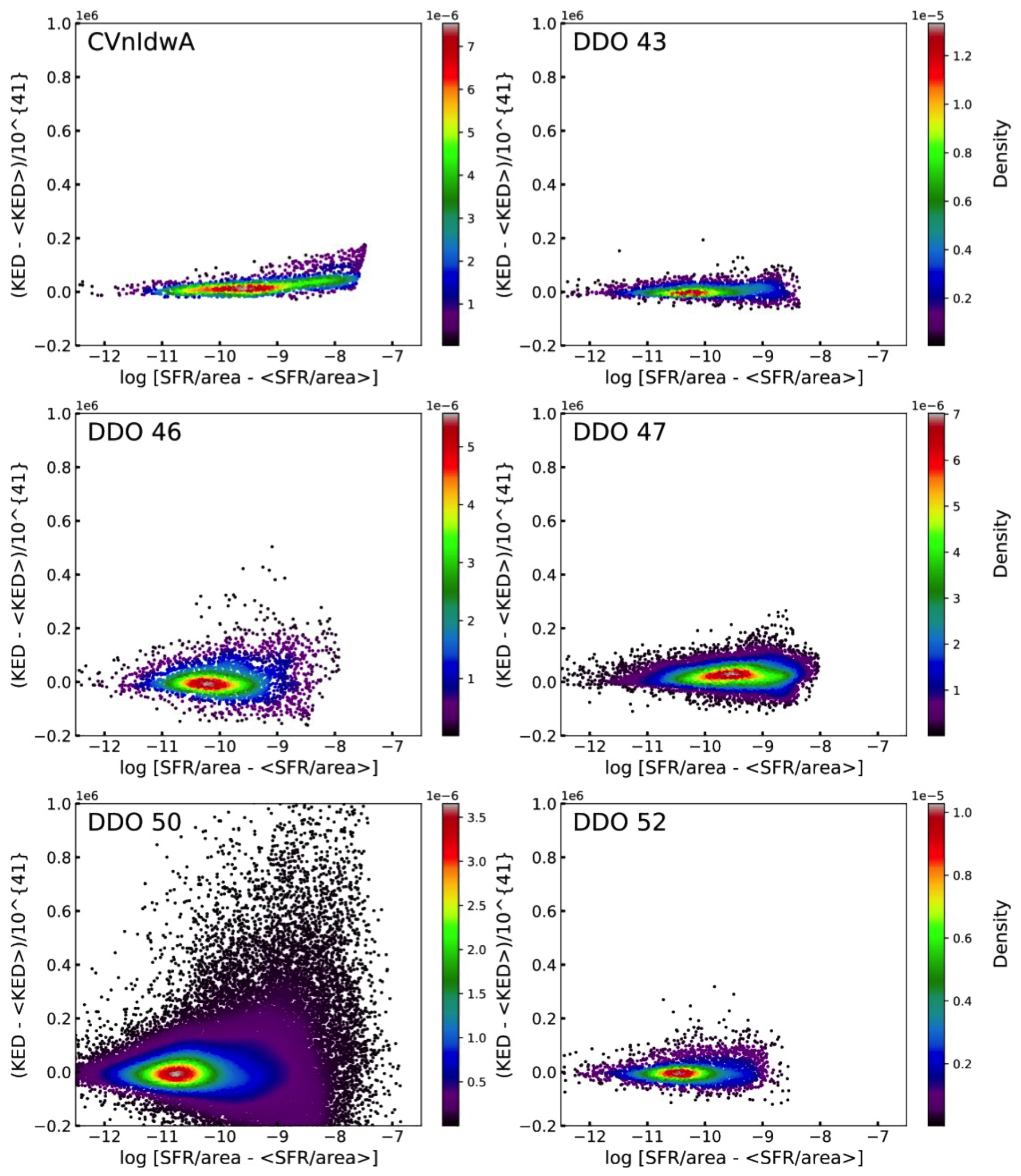}
\vskip -0.1truein
\caption{
Pixel-pixel plots of the excess KED above the average value at each radius vs.\
the log of the excess \sfra. The density of points is color-coded.
Figures for the rest of the galaxies in this study are available in the on-line materials
(there are 6 figures like this for 36 galaxies).  The KED has not been
corrected for He and heavy elements.
\label{fig-pixpixKED}}
\end{figure}

\begin{figure}[t!]
\epsscale{1.0}
\vskip -0.7truein
\plotone{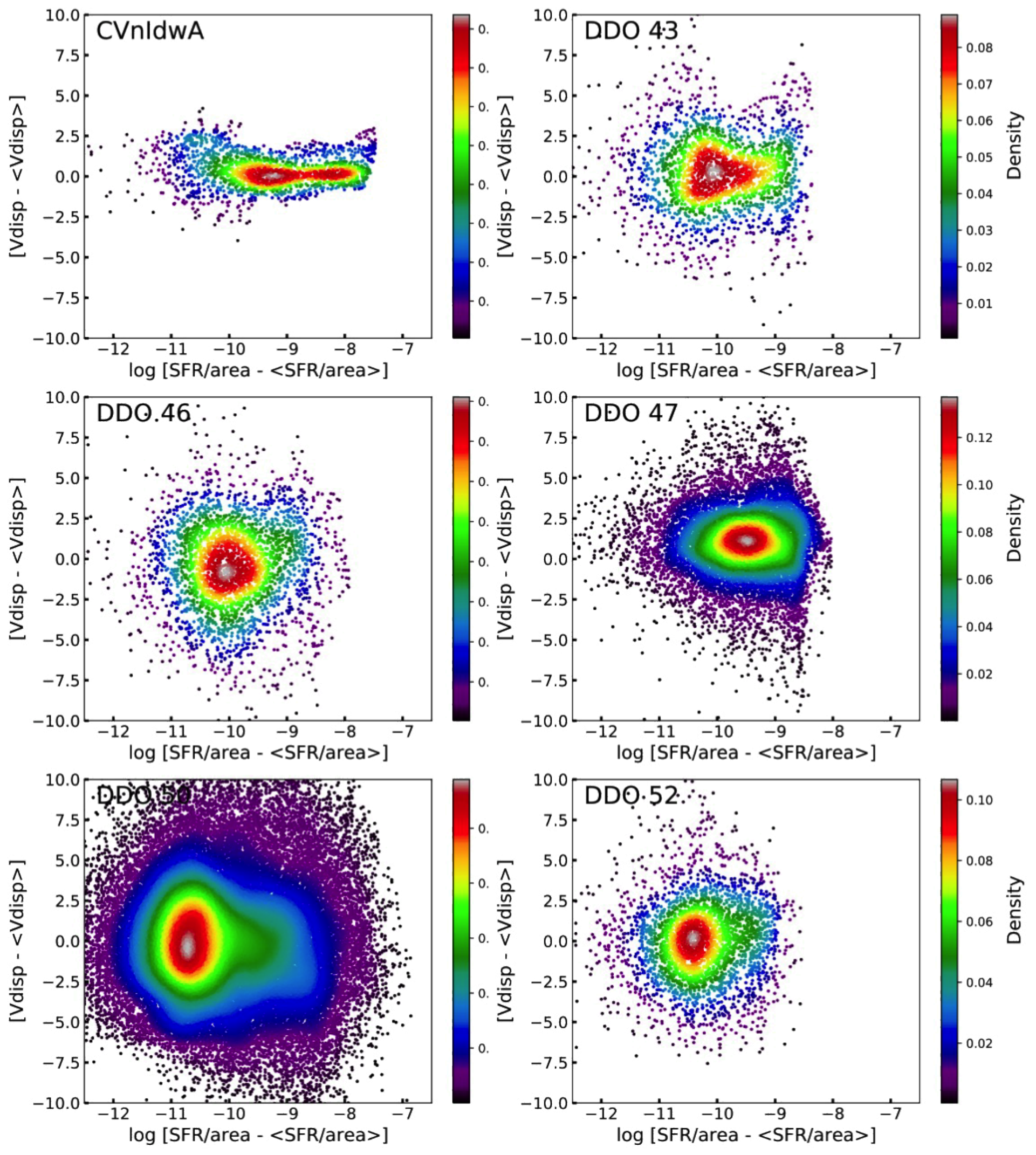}
\vskip -0.2truein
\caption{
Pixel-pixel plots of the excess \vdisp\ above the average value at each radius
vs.\ the log of the excess \sfra. The density of points is color-coded.
Figures for the rest of the galaxies in this study are available in the on-line materials
(the remaining 36 galaxies are shown in 6 figures).
\label{fig-pixpixVdisp}}
\end{figure}

\begin{figure}[t!]
\epsscale{1.0}
\vskip -0.7truein
\plotone{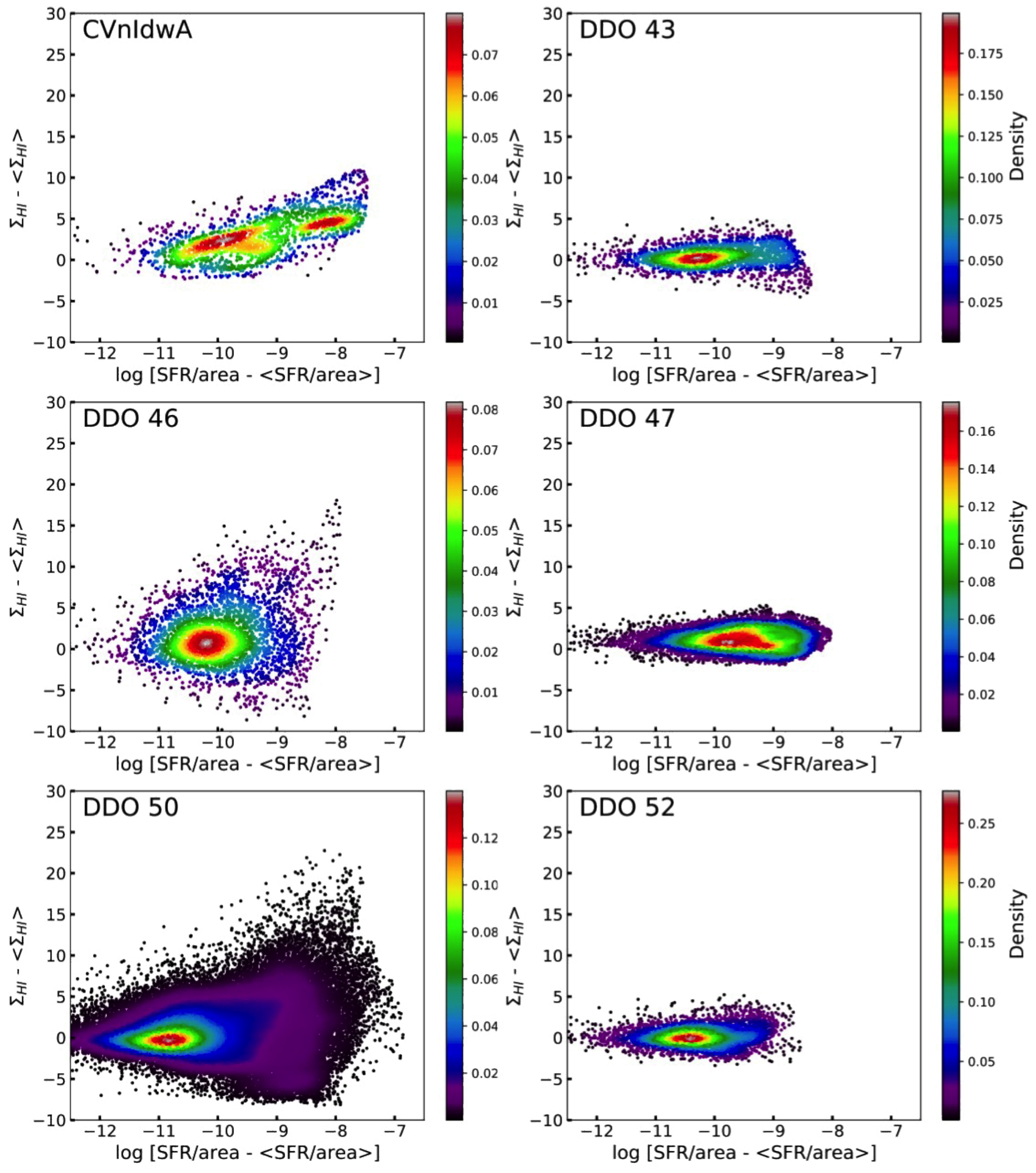}
\vskip -0.2truein
\caption{
Pixel-pixel plots of the excess \sighi\ above the average value at each radius
vs.\ log of the excess \sfra. The density of points is color-coded.
Figures for the rest of the galaxies in this study are available in the on-line materials
(the remaining 36 galaxies are shown in 6 figures).
\label{fig-pixpixsighi}}
\end{figure}

Another way of looking at the data is to compare individual pixels in pairs of
images. We have done that, examining KED, the velocity dispersion of the gas
\vdisp, and \HI\ surface density \sighi\ versus SFR surface density as determined
from the FUV images, \sfra. Recall that the FUV images were geometrically
transformed and smoothed to match the pixel size and resolution of the \HI\
images. For all galaxies but DDO 216 and Sag DIG, the pixel size is
$1.5^{\prime\prime}$ and for these two it is $3.5^{\prime\prime}$. To compensate
for radial trends, we determined the azimuthally-averaged \sfra, KED, \HI\
and \vdisp\ in annuli from the center of the galaxy and subtracted that from the
observations.  We used optically-determined disk parameters of center,
minor-to-major axis ratio $b/a$, and position angle of the major axis from
\citet{lt12}. The widths of the annuli, constant in a given galaxy, were
chosen to be the same as those used to measure the \HI\ surface density profiles
of \citet{lt12}. The azimuthally-averaged radial profiles of \sfra, KED, \vdisp,
and \sighi\ are shown for each galaxy in Figure \ref{fig-radialprof}. The
pixel-pixel plots of excess KED, \vdisp\ and \sighi\ versus excess \sfra\ are
shown in Figures \ref{fig-pixpixKED}-\ref{fig-pixpixsighi}. All of these
quantities except \vdisp\ were corrected to a face-on orientation by multiplying
the fluxes by the cosine of the inclination. The KED units are erg pc$^{-2}$,
\vdisp\ is in km s$^{-1}$, \sighi\ is in M\solar pc$^{-2}$ and \sfra\ is in units
of $M_\odot$ yr$^{-1}$ pc$^{-2}$. KED values in Figures
\ref{fig-radialprof}-\ref{fig-pixpixsighi} have not been corrected for Helium and
heavy elements. Note that only the regions of relatively high \sfra\ are plotted,
i.e., with positive excess above the annular average, and we plot the logarithm of
this excess. For the quantities on the ordinate, we consider both positive and
negative excess values over the average, so they are not plotted in the log. Some
regions of locally high \sfra\ have locally low KED, \vdisp\ or \sighi.

In the radial averages shown in Figure \ref{fig-radialprof}, we see that KED, \sfra, and \sighi\ generally decline with radius.
\citet{tamburro09} found this also for spiral galaxies. They also found that \vdisp\ declines with radius in their sample, but in our sample
of \dirr\ the drop of \vdisp\ with radius is very minor, if any. They also find a clear correlation of KED with \sfra\ in pixel-pixel plots,
whereas our Figure \ref{fig-pixpixKED} does not show such a nice correlation.

\clearpage

\hspace{-0.7truein} \vspace{-0.35truein}
\includegraphics[scale=1.0]{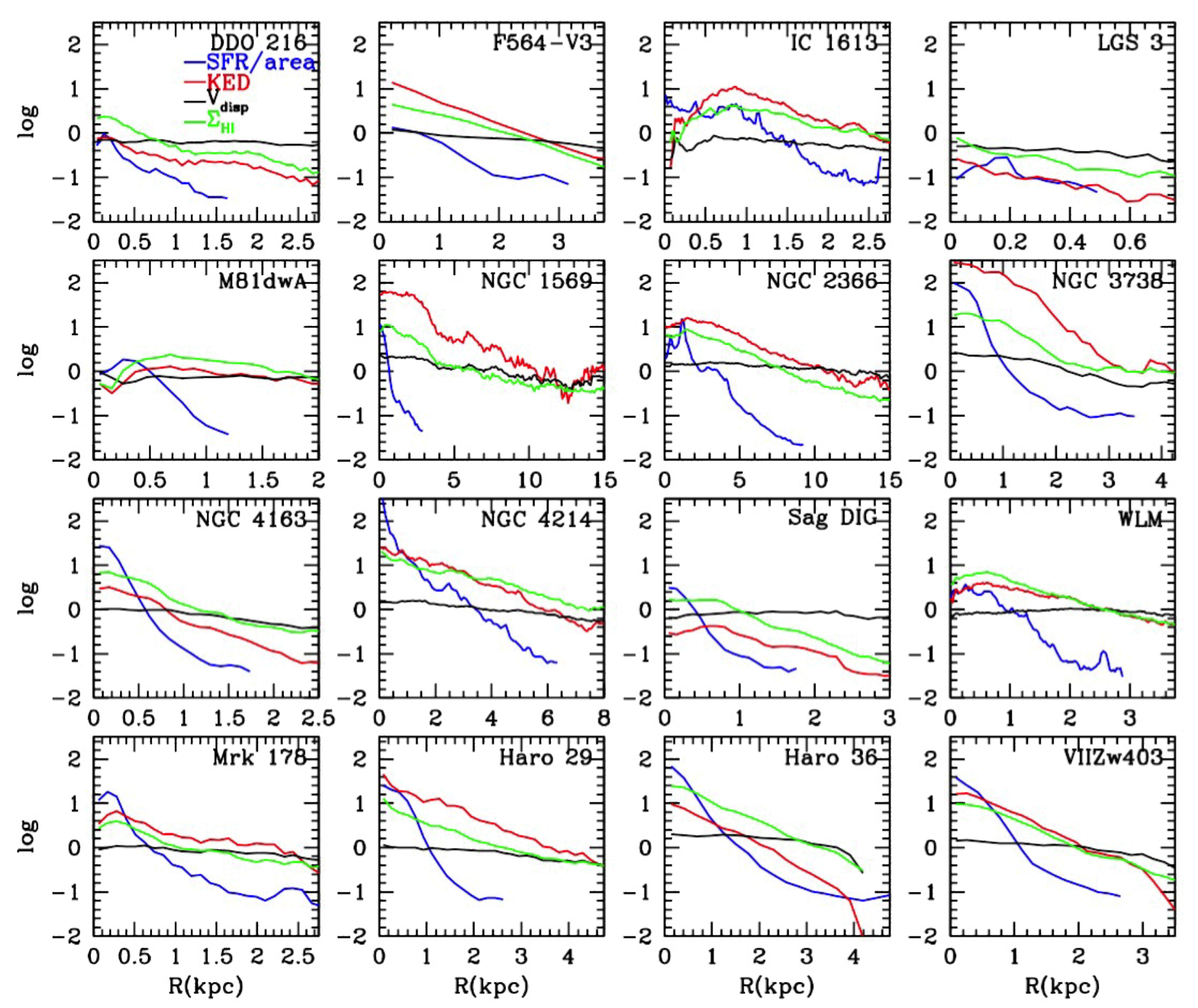}

\clearpage

\clearpage

\clearpage

Figures \ref{fig-pixpixKED} - \ref{fig-pixpixsighi} typically show concentrations
of points at a low excess \sfra\ and a continuation of these points toward higher
excess \sfra. The low excess \sfra\ are in the outer disks and the high excess
\sfra\ are in the inner disks. Some galaxies have two concentrations of points in
these figures.

To quantify the pixel distributions, we determined the excess \sfra\ and other
quantities at the plotted concentrations. For each galaxy we made a histogram of
the log of the excess \sfra\ (the abscissa value) and found the peak at the low
density concentration.  The excess log \sfra\ in that peak was determined from the
average value in the three bins of the histogram centered there. The bin width was
0.2 in the log of the excess \sfra. Then for these three bins around the histogram
peak for log excess \sfra, we determined the mean value of the quantity plotted on
the ordinate in the figures, i.e., the excess KED, \vdisp\, and \sighi. For the
higher excess \sfra, we took the mean value of the excess \sfra\ and the other
quantities for all regions where the log of the excess \sfra\ was larger than the
high-SF edge of the concentration of points, typically at $-9.8$ but ranging from
$-9.5$ to $-10.2$ depending on the plotted galaxy. When there was only one
prominent concentration of points in the figure, we determined the values there.

\begin{figure}[t!]
\epsscale{1.0}
\vskip -0.4truein
\plotone{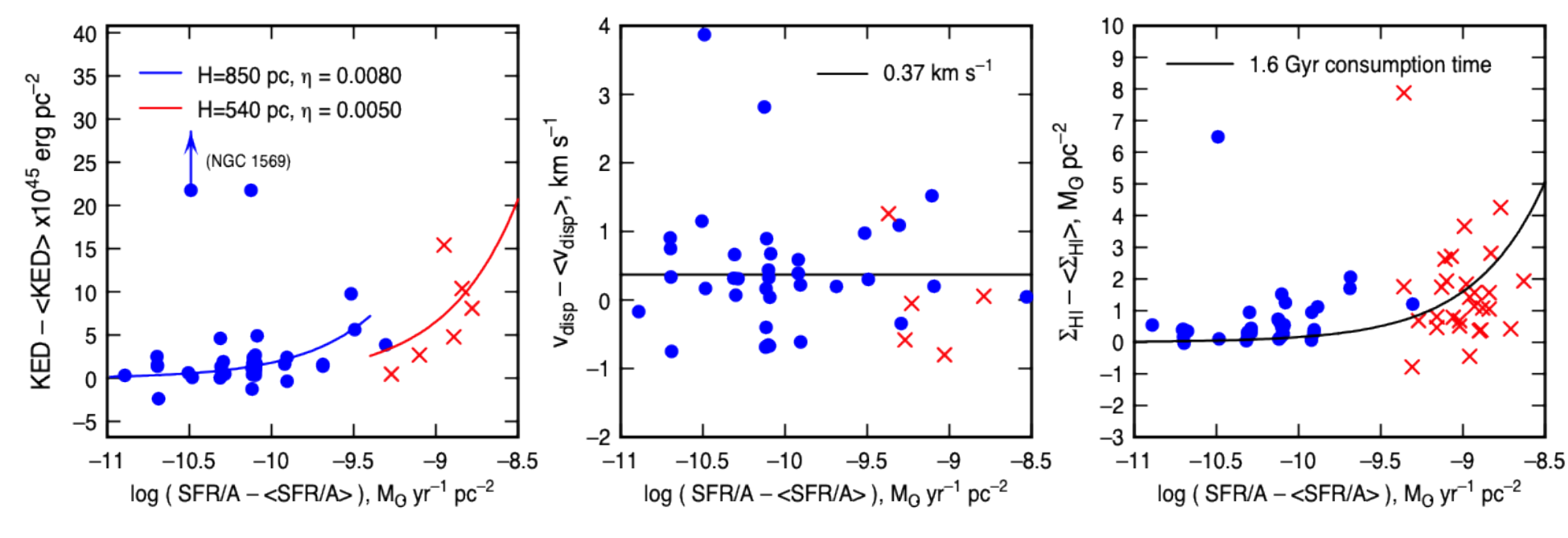}
\vskip -0.25truein
\caption{Mean excess KED corrected for He and heavy elements,
\vdisp, and \sighi\ versus the mean of the log of the
(positive) excess \sfra\ for all galaxies.
Dots correspond to the low \sfra\ concentrations in the outer disks
and crosses correspond to high \sfra\ in the inner disks.
{\it Left:}  The curves show fitted relationships between the KED generated by supernovae
and the \sfra\ for the indicated efficiencies of converting SN energy into
turbulence, and for galaxy scale heights of 850 pc and 540 pc.
{\it Middle:} Average \vdisp\ excess at each concentration of excess \sfra\ in Figure \ref{fig-pixpixVdisp}.
The excess velocity dispersion averages 0.34 km s$^{-1}$ in the outer disk
and 0.17 km s$^{-1}$ in the inner disk.
{\it Right:} Average \sighi\ excess at each concentration of excess \sfra\ in Figure \ref{fig-pixpixsighi}.
There is a clear trend toward excess \HI\ at local SF regions.
\label{fig-pixel_correlations}}
\end{figure}

Figure \ref{fig-pixel_correlations} shows the mean excess KED corrected for
Helium and heavy elements, \vdisp, and \sighi\ versus the mean of the log of the
(positive) excess \sfra\ for all galaxies, with dots corresponding to the low
\sfra\ concentrations in the outer disks and crosses corresponding to the high
\sfra\ in the inner disks. The curves in the KED plot show fitted relationships
between the KED generated by supernovae and the \sfra\ for the indicated
efficiencies of converting SN energy into turbulence, and for galaxy scale heights
of 850 pc and 540 pc. These theoretical KEDs come from equation 3.7 in
\cite{bacchini20}, which is
\begin{equation}
KED_{\rm SN} = \eta\Sigma_{\rm SFR}f_{\rm cc}E_{\rm SN}(2H/v_{\rm turb})
\label{eq:bacchini}
\end{equation}
where $\eta$ is the efficiency of energy conversion from supernova to turbulence,
$f_{\rm cc}=1.3\times10^{-2}\;M_\odot^{-1}$ is the number of core-collapse
supernovae per solar mass of stars, $E_{\rm SN}=10^{51}$ erg is the supernova
energy, $H$ is the disk thickness and $v_{\rm turb}$ is the turbulent gas velocity
dispersion (the ratio of these latter two quantities gives the turbulent
dissipation time). \cite{bacchini20} compare the radial profiles of turbulent
energies in 10 nearby galaxies with the SFRs and derive an average efficiency of
$1.5^{1.8}_{0.8}$\% if all of the turbulence comes from star formation. Because
the required efficiency is relatively low, they concluded that supernovae related
to star formation can drive most of the interstellar turbulence.

For the dwarf galaxies studied here, we evaluate equation (\ref{eq:bacchini})
using scale heights and velocity dispersions from the average values for 20 dIrrs
in \cite{eh15}, in Table 2 of that paper. For the concentrations of pixel values
corresponding to the outer regions of the galaxies, we take the average scale
height and \vdisp\ at 2 scale lengths, which are $H=850$ pc and $v_{\rm disp}=9.7$
km s$^{-1}$. For the inner regions, we take the values at 1 scale length, which
are 540 pc and 10.7 km s$^{-1}$. We also include Helium and heavy elements in
the KED by multiplying the \HI\ mass surface density by 1.36. Then with $f_{\rm
cc}$ and $E_{\rm SN}$ given above, equation (\ref{eq:bacchini}) is fitted for the
efficiency in the two cases, using for the moment $v_{\rm disp}$ instead of
$v_{\rm turb}$. The results are drawn as curves in the left panel of Figure
\ref{fig-pixel_correlations}. The average local efficiencies for conversion of SN
energy to KED are $\eta=0.0080\pm0.045$ and $0.0050\pm0.0075$ for the outer
and inner disk regions, with these assumptions.

The total dispersions used to evaluate $\eta$ include thermal and turbulent
motions, which were distinguished in several limiting cases by \cite{bacchini20}
to get the desired $v_{\rm turb}$. If we assume Mach$\sim1$ turbulence in the
general \HI\ ISM, then $v_{\rm turb}=v_{\rm disp}/2^{0.5}$ and the derived values
of $\eta$ decrease by the factor 0.7, preserving the ratio $\eta/v$ used to match
the KED. Alternatively, we could use thermal dispersions of 4.9 km s$^{-1}$ and
6.1 km s$^{-1}$ modeled for NGC 4736 and NGC 2403 respectively by
\cite{bacchini20} to estimate that $v_{\rm turb}/v_{\rm disp}\sim0.8$, given that
$v_{\rm disp}\sim10$ km s$^{-1}$ here. Then our derived $\eta$ should decrease by
$\sim0.8$.  The \cite{bacchini20} galaxies were more massive than our dIrr
galaxies, but the thermal contributions to $v_{\rm disp}$ are not likely to be
much different. These corrections change the average value of $\eta=0.0065$ for
the two regions in Figure \ref{fig-pixel_correlations} to $\eta\sim0.0048$, using
a mean correction factor of 0.75.

This $\eta$ value is the average for the peak regions of star formation. It
measures how efficiently star formation energy gets into \HI\ turbulent motions
locally in units of the supernova energy per unit mass of young stars.  When
normalized this way, other types of energy related to star formation such as
expanding HII regions and stellar winds are included in $\eta$ too. What is not
included as a source of turbulent motion is energy unrelated to star formation,
such as gravitational energy from gas collapse on the scale of the ISM Jeans
length, or collapse energy from transient spiral arms driven by combined gas and
stellar masses, or shock energy from the relative motions of gas and stellar
spiral density waves. If $\eta\sim0.0048$ measured locally gives the actual
efficiency for star formation to pump turbulence in the \HI\ gas, then the global
turbulent energy pumped by all of the star formation in a galaxy should equal our
local $\eta$ multiplied by the global star formation rate (along with the other
factors in equation \ref{eq:bacchini}). Because \cite{bacchini20} found that the
global turbulent energy is $1.5^{1.8}_{0.8}$\% of the energy derived from the star
formation rate, there would seem to be more energy required than what star
formation alone can provide. The excess energy needed is a factor of
$\sim0.015/0.0048-1\sim2$ times the star formation energy.

This factor has many uncertainties, both from the range in global values
derived by \cite{bacchini20} and from the galaxy-to-galaxy or inner-disk to
outer-disk variations derived here. For example, our $\eta\sim0.0048$ is closer to
that of the dwarf galaxy DDO154 in \cite{bacchini20}, which had $\eta=0.009$
assuming a pure warm phase \HI. Also, our average $\eta$ for the inner disk
regions in Figure \ref{fig-pixel_correlations} was higher than the average for the
inner and outer disks combined (which gave the value 0.0048) by a factor of 1.2.
But even within this range, the global energy from turbulence seems to be larger
than what can be pumped from star formation alone, if we use local star formation
rates as the basic means of calibrating $\eta$.

Figure \ref{fig-pixel_correlations} for the KED excess has two high points for the
inner disk which were not included in the efficiency fit. These are the galaxies
Haro 29 with excess KED$=16.0\times10^{45}$ erg pc$^{-2}$, and NGC 1569 with excess
KED$=141\times10^{45}$ erg pc$^{-2}$. Correspondingly, Figure \ref{fig-pixpixKED}
shows a scatter of individual pixel points to very high values of KED for these
galaxies.

The middle panel of Figure \ref{fig-pixel_correlations} shows the average \vdisp\
excess at each concentration of excess \sfra\ in Figure \ref{fig-pixpixVdisp}. The
excess velocity dispersion is rarely larger than 1 km s$^{-1}$ and averages 0.45
km s$^{-1}$ in the outer disk, $-0.34$ km s$^{-1}$ in the inner disk, and 0.37 km
s$^{-1}$ overall. Some local star formation regions have lower \HI\ velocity
dispersions than the average at that galactocentric radius, giving negative
excesses in Figure \ref{fig-pixel_correlations}.  These typically small excesses
in the local \HI\ velocity dispersion are consistent with the small feedback
efficiencies found above. There is relatively little generation of turbulence at
the positions of star-forming regions.

The right-hand panel of Figure \ref{fig-pixel_correlations} shows the average
\sighi\ excess at each concentration of excess \sfra\ in Figure
\ref{fig-pixpixsighi}. There is a clear trend toward excess \HI\ at local SF
regions, although in a few cases the \HI\ is less than the azimuthal average. This
general excess corresponds to a ratio of \sighi\ to \sfra\ that equals 6.5 Gyr in
the outer disk, 1.2 Gyr in the inner disk and 1.6 Gyr overall, where this latter
fit is shown by the curve in the figure. For this fit, the high point that is
plotted in Figure \ref{fig-pixel_correlations} is excluded; that is for NGC 1569,
where the ratio is 31 Gyr. This average ratio of $\sim1.6$ Gyr is comparable to the
consumption time for molecules, which is about 2 Gyr in \cite{bigiel08} and
\cite{leroy08}. If only molecular clouds form stars, then this similarity
implies that the molecular fraction is about 50\%
in the inner disk, as suggested using other properties of \HI\ and star formation
in recent papers \citep{hunter19,hunter20,madden20}.

\section{Discussion}

Comparisons between the kinetic energy density or velocity dispersion and the local
star formation rate using cross correlations of several types and pixel-level
excesses above the radial average quantities have shown virtually no connections
between large-scale turbulence and star formation. Many of the galaxies have lumpy
KED and FUV images but the lumps are not well correlated or anti-correlated spatially. This is
contrary to some theoretical expectations and the simulations that have been
designed to illustrate those expectations which suggest that feedback from star
formation pumps a significant amount of interstellar turbulence, and thereby
controls the interstellar scale height and average mid-plane density. While it is
generally accepted that this mid-plane density controls the collapse rate of the
ISM and therefore the average star formation rate, the origin of the turbulence and
other vertical forces which determine the scale height and density have been
difficult to observe directly. Most likely, the maintenance of a modest value for
gravitational stability parameter $Q$ controls the overall interstellar turbulent
speed through pervasive and mild gravitational instabilities, which also feed the
star formation process through cloud formation. This was demonstrated by
\cite{bournaud10} and also underlies the Feedback in Realistic Environments (FIRE)
simulations by \citet{FIRE1}; the primary role of feedback is to destroy molecular
clouds locally \citep{FIRE2}. Our data suggest that this feedback does not extend
far enough from molecular clouds to be visible in the \HI\ at our resolution
(from 26 pc at IC 1613 to 340 pc at DDO 52).

\section{Summary} \label{sec-summary}

We have examined the relationship between star formation, as traced by FUV images,
and turbulence in the gas, as traced by kinetic energy density images and velocity
dispersion maps in the LITTLE THINGS sample of nearby \dirr\ galaxies. We performed
2D cross-correlations between FUV and KED images, finding maximum \cc\ that
indicate little correlation. A plot of integrated SFR against the maximum
\cc\ also shows no correlation. We also performed cross-correlations in annuli
centered on the optical center of the galaxy to produce \cc\ as a function of
radius. In some galaxies the centers have \cc\ that are high enough to indicate a
correlation, and in some galaxies the \cc\ drops off with radius from the center.

To look at the images a different way, we determined the fraction of pixels in the
FUV and KED images with values above a given percentage of the maximum pixel value
in the image. Plots of these quantities show different behaviors for FUV and KED
images in many of the galaxies.  Finally, we considered  on a pixel-by-pixel basis
the excess KED, \vdisp, and \sighi\ above the average radial profiles of these
quantities and plotted that versus the excess \sfra. There was a weak tendency to
have a higher excess KED at a higher excess \sfra, corresponding to an efficiency
of kinetic energy input to the local ISM from supernova related to star formation
of about 0.5\%. This is too small by a factor of about 2 for star
formation to be the only source of global kinetic energy density. The excess
\vdisp\ connected with star formation peaks is also small, only $0.37$ km s$^{-1}$
on average.  The angular scale for these small excesses is typically
$1.5^{\prime\prime}$, which, for a distance of 3 Mpc, corresponds to $\sim20$ pc.

\acknowledgments

We are grateful to Dr. C. Bacchini for comments on the manuscript.  H.A.\ is
grateful to the Lowell Observatory Director's Opportunity Network for funding to
work on this project. Lowell Observatory sits at the base of mountains sacred to
tribes throughout the region. We honor their past, present, and future
generations, who have lived here for millennia and will forever call this place
home.

Facilities: \facility{VLA} \facility{GALEX}


\begin{thebibliography}{}
\bibitem[Bacchini et al.(2020)]{bacchini20} Bacchini, C., Fraternali, F., Iorio, G., Pezzulli, G. Marasco, A. \& Nipoti, C. 2020, arXiv:2006.10764
\bibitem[Benincasa et al.(2020)]{FIRE2} Benincasa, S.\ M., Loebman, S.\ R., Wetzel, A., et al. 2020, \mnras, submitted, arXiv:1911.05251
\bibitem[Bigiel et al.(2008)]{bigiel08} Bigiel, F., Leroy, A., Walter, F., et al. 2008, \aj, 136, 2846
\bibitem[Bigiel et al.(2010)]{bigiel10} Bigiel, F., Leroy, A., Walter, F., et al. 2010, \aj, 140, 1194
\bibitem[Bournaud et al.(2010)]{bournaud10} Bournaud, F., Elmegreen, B.G., Teyssier, R., Block, D.L., \& Puerari, I. 2010, MNRAS, 409, 1088
\bibitem[Burkhart et al.(2010)]{burkhart10} Burkhart, B., Stanimirovi\'{c}, S., Lazarian, A., \& Kowal, G. 2010, \apj, 708, 1204
\bibitem[Combes et al.(2012)]{combes12} Combes, F., Boquien, M., Kramer, C. et al. 2012, A\&A, 539, A67
\bibitem[Deharveng et al.(2012)]{deharveng12} Deharveng, L., Zavagno, A., Anderson, L. D., Motte, F., Abergel, A., Andr\'e, Ph. Bontemps, S., Leleu, G., Roussel, H., Russeil, D. 2012, A\&A, 546, A74
\bibitem[Dib \& Burkert(2005)]{dib05} Dib, S., \& Burkert, A. 2005, \apj, 630, 238
\bibitem[Efremov \& Elmegreen(1998)]{efremov98} Efremov, Y.\ N., \& Elmegreen, B.\ G. 1998, \mnras, 299, 588
\bibitem[Egorov et al.(2017)]{egorov17} Egorov, O.V., Lozinskaya, T.A., Moiseev, A.V., \& Shchekinov, Y.A. 2017, MNRAS, 464, 1833
\bibitem[{Elmegreen}(1993)]{elmegreen93} Elmegreen, B.G. 1993, \apj, 419, L29
\bibitem[{Elmegreen \& Efremov}(1997)]{elmegreen97} Elmegreen, B.\ G., \& Efremov, Y.\ N. 1997, \apj, 480, 235
\bibitem[Elmegreen \& Hunter(2006)]{eh06} Elmegreen, B.\ G., \& Hunter, D.\ A. 2006, \apj, 636, 712
\bibitem[Elmegreen \& Hunter(2015)]{eh15} Elmegreen, B.\ G., \& Hunter, D.\ A. 2015, \apj, 805, 145
\bibitem[Goldbaum et al.(2016)]{goldbaum16} Goldbaum, N.J., Krumholz, M.R., Forbes, J.C. 2016, ApJ, 827, 28
\bibitem[Herrmann et al.(2013)]{herrmann13} Herrmann, K.\ A., Hunter, D.\ A., \& Elmegreen, B. G.\ 2013, \aj, 146, 104
\bibitem[Hopkins et al.(2011)]{hopkins11} Hopkins, Philip F., Quataert, E., \& Murray, N.  2011, MNRAS, 417, 950
\bibitem[Hopkins et al.(2014)]{FIRE1} Hopkins, P.\ F., Kere\u{s}, D., O\~{n}orbe, J., et al. 2014, \mnras, 445, 581
\bibitem[Hunter \& Elmegreen(2004)]{he04} Hunter, D.\ A., \& Elmegreen, B.\ G. 2004, \aj, 128, 2170
\bibitem[Hunter et al.(2003)]{hunter03} Hunter, D.\ A., Elmegreen, B.\ G., Dupuy, T.\ J., \& Mortonson, M. 2003, \aj, 126, 1836
\bibitem[Hunter et al.(2016)]{outerfuv} Hunter, D.\ A., Elmegreen, B.\ G., \& Gehret, E.\ 2016, \aj, 151, 136
\bibitem[Hunter et al.(2019)]{hunter19} Hunter, D.\ A., Elmegreen, B.\ G., \& Berger, C.\ L. 2019, \aj, 157, 241
\bibitem[Hunter et al.(2020)]{hunter20} Hunter, D.A., Elmegreen, B.G., Goldberger, E., et al. 2020, \aj, submitted
\bibitem[Hunter et al.(2010)]{hunter10} Hunter, D.\ A., Elmegreen, B.\ G. \& Ludka, B.\ C. 2010, \aj, 139, 447
\bibitem[Hunter et al.(2011)]{hunter11} Hunter, D.\ A., Elmegreen, B.\ G., Oh, S.-H., et al. 2011,\aj, 142, 121
\bibitem[Hunter et al.(2013)]{spirals} Hunter, D.\ A., Elmegreen, B.\ G., Rubin, V.\ C., \& Ashburn, A. 2013, \aj, 146, 92
\bibitem[Hunter et al.(2001)]{hunter01} Hunter, D.\ A., Elmegreen, B.\ G., \& van Woerden, H.  2001, \apj, 556, 773
\bibitem[Hunter et al.(2012)]{lt12} Hunter, D.\ A., Ficut-Vicas, D., Ashley, T., et al.\ 2012, \aj, 144, id 13
\bibitem[Hunter \& Plummer(1996)]{sextansa} Hunter, D.\ A., \& Plummer, J.\ D. 1996, \apj, 462, 732
\bibitem[Ib\'a\~nez-Mej\'ia et al.(2017)]{ibanez17} Ib\'a\~nez-Mej\'ia, J.C., Mac Low, M.-M., Klessen, R.S., Baczynski, C. 2017, ApJ, 850, 62
\bibitem[Joung et al.(2009)]{joung09} Joung, M.R., Mac Low, M.-M., \& Bryan, G.L. 2009, ApJ, 704, 137
\bibitem[Kennicutt(1989)]{kennicutt89} Kennicutt, R.\ C., Jr. 1989, \apj, 344, 685
\bibitem[Kim \& Ostriker(2015)]{kim15} Kim, C.-G., \& Ostriker, E. C. 2015, ApJ, 815, 67
\bibitem[Kim et al.(2018)]{kim18} Kim, J.-G., Kim, W.-T., Ostriker, E.C. 2018, ApJ, 859, 68
\bibitem[Kingsburgh \& McCall(1998)]{kingsburgh98} Kingsburgh, R.\ L., \& McCall, M.\ L. 1998, \aj, 116, 2246
\bibitem[Kraljic et al.(2014)]{kraljic14} Kraljic, K., Renaud, F., Bournaud, F., et al. 2014, \apj, 784, 112
\bibitem[Krumholz \& McKee(2005)]{krumholz05} Krumholz, M.\ R., \& McKee, C.\ F. 2005, \apj, 630, 250
\bibitem[Krumholz et al.(2018)]{krumholz18} Krumholz, M.R., Burkhart, B., Forbes, J.C., \& Crocker, R.M. 2018, MNRAS, 477, 2716
\bibitem[Lehnert et al.(2013)]{lehnert13} Lehnert M. D., Le Tiran L., Nesvadba N. P. H., van Driel W., Boulanger F., Di Matteo P., 2013, A\&A, 555, A72
\bibitem[Leroy et al.(2008)]{leroy08} Leroy, A.\ K., Walter, F., Brinks, E., Bigiel, F., de Blok, W.\ J.\ G., Madore, B., \& Thornley, M.\ D., 2008, AJ, 136, 2782
\bibitem[Mac Low et al.(2017)]{maclow17} Mac Low, M.-M., Burkert, A., \& Ib\'{a}\~{n}ez-Mej\'{i}a, J. C. 2017, ApJL, 847, L10
\bibitem[Mac Low \& Klessen(2004)]{maclow04} Mac Low, M.-M., \& Klessen, R.\ S. 2004, Rev Mod Phys, 76, 125
\bibitem[Madden et al.(2020)]{madden20} Madden, S.\ C., Cormier, D., Hony, S., Lebouteiller, V., Abel, N., et al. 2020, A\&A, 643, A141
\bibitem[Maier et al.(2016)]{maier16} Maier, E., Chien, L.-H., \& Hunter, D.\ A. \aj, 152, 134
\bibitem[Maier et al.(2017)]{maier17} Maier, E., Elmegreen, B.\ G., Hunter, D.\ A., et al.\ 2017, \aj, 153, 163
\bibitem[Martin et al.(2005)]{galex} Martin, D.\ C., Fanson, J., Schiminovich, D., et al.\ 2005, \apj, 619, L1
\bibitem[Meurer et al.(1996)]{meurer96} Meurer, G.\ R., Carignan, C., Beaulieu, S.\ F., \& Freeman, K.\ C. 1996, \aj, 111, 1551
\bibitem[Moiseev et al.(2015)]{moiseev15} Moiseev, A.\ V., Tikhonov, A.\ V., \& Klypin, A. 2015, \mnras, 449, 3568
\bibitem[Ossenkopf et al.(2008)]{ossenkopf08} Ossenkopf, V., Krips, M., \& Stutzki, J. 2008, \aa, 485, 917
\bibitem[Ostriker et al.(2010)]{ostriker10} Ostriker, E.\ C., McKee, C.\ F., \& Leroy, A.\ K. 2010, \apj, 721, 975
\bibitem[Padoan et al.(2016)]{padoan16} Padoan, P., Pan, L., Haugb\/olle, T., \& Nordlund, \AA. 2016, ApJ, 822, 11
\bibitem[Palmeirim et al.(2017)]{palmeirim17} Palmeirim, P., Zavagno, A., Elia, D. et al. 2017, A\&A, 605, A35
\bibitem[Piontek \& Ostriker(2005)]{piontek05} Piontek, R.\ A., \& Ostriker, E.\ C. 2005,  \apj, 629, 849
\bibitem[Romeo \& Mogotsi(2017)]{romeo17} Romeo, A.B., \& Mogotsi, K.M. 2017, MNRAS, 469, 286
\bibitem[Stilp et al.(2013)]{stilp13} Stilp, A. M., Dalcanton, J. J., Skillman, E., Warren, S. R., Ott, J., \& Koribalski, B. 2013, ApJ, 733, 88
\bibitem[Struck \& Smith(1999)]{struck99} Struck, C., \& Smith, D.\ C. 1999, \apj, 527, 673
\bibitem[Tamburro et al.(2009)]{tamburro09} Tamburro, D., Rix, H.-W., Leroy, A.\ K., Mac Low, M.-M., Walter, F., et al. \aj, 137, 4424
\bibitem[Toomre(1964)]{toomre} Toomre, A. 1964, \apj, 139, 1217
\bibitem[\"Ubler et al.(2019)]{ubler19} \"Ubler, H., Genzel, R., Wisnioski, E. et al. 2019, ApJ, 880, 48
\bibitem[van Zee et al.(1997)]{vanzee97} van Zee, L., Haynes, M.\ P., Salzer, J.\ J., \& Broeils, A.\ H. 1997, \aj, 113, 1618
\bibitem[Varidel et al.(2020)]{varidel20} Varidel, M.R., Croom, S.M., Lewis, G.F. et al. 2020, MNRAS, 495, 2265
\bibitem[Walter et al.(2008)]{walter08} Walter, F., Brinks, E., de Blok, W.~J.~G., Bigiel, F., Kennicutt, R.~C., Jr., et al.  2008, \aj, 136, 2563
\bibitem[Willett et al.(2005)]{willett05}  Willett, K.\ W., Elmegreen, B.\ G., \& Hunter, D.\ A. 2005, \aj, 129, 2186
\bibitem[Youngblood \& Hunter(1999)]{youngblood99} Youngblood, A.\ J., \& Hunter, D.\ A. 1999, \apj, 519, 55
\bibitem[Zhang et al.(2012)]{zhang12} Zhang, H.-X., Hunter, D.\ A., \& Elmegreen, B.\ G. 2012, \apj, 754, 29
\bibitem[Zhou et al.(2017)]{zhou17} Zhou, L., Federrath, C., Yuan, T., et al. 2017, MNRAS, 470, 4573
\end{thebibliography}
\end{document}